\documentclass{article}
\pdfoutput=1
\usepackage{lscape}
\usepackage{enumerate}
\usepackage{amsfonts}
\usepackage{amsmath}
\usepackage{textcomp}
\usepackage{amssymb}
\usepackage{color}
\usepackage{graphicx}
\usepackage{pdflscape}
\usepackage{afterpage}
\usepackage[matrix,arrow,curve]{xy}
\usepackage{array}
\usepackage{indentfirst}
\usepackage{subcaption}

\def\be{\begin{equation}}
\def\ee{\end{equation}}

\makeatletter
\renewcommand*{\@cite@ofmt}{\bfseries\hbox}
\makeatother

\def\green{\color{green}}
\def\blue{\color{blue}}
\def\red{\color{red}}
\def\black{\color{black}}

\hoffset= - 1.2in         

\begin{document}

\title{\vspace{0.1cm}{\Large {\bf  Entangling gates for the $SU(N)$ anyons}\vspace{.2cm}}
\author{\bf Sergey Mironov$^{a,c,d}$\thanks{e-mail: sa.mironov\_1@physics.msu.ru},
Andrey Morozov$^{b,c,e}$\thanks{e-mail: morozov.andrey.a@iitp.ru}}
}
\date{ }

\maketitle

\vspace{-5.5cm}

\begin{center}
\hfill ITEP/TH-15/26\\
\hfill IITP/TH-14/26\\
\hfill MIPT/TH-14/26\\
\end{center}

\vspace{3.6cm}

\begin{center}
$^a$ {\small {\it INR RAS, Moscow 117312, Russia}}\\
$^b$ {\small {\it IITP RAS, Moscow 127994, Russia}}\\
$^c$ {\small {\it NRC ``Kurchatov Institute'', Moscow 123182, Russia}}\\
$^d$ {\small {\it ITMP, MSU, Moscow, 119991, Russia}}\\
$^e$ {\small {\it MIPT, Dolgoprudny, 141701, Russia}}\\
\end{center}

\vspace{1cm}

\begin{abstract}
The model of a topological quantum computer is a promising one due to its natural resistance to noise and other errors. Operations in such a computer are implemented by braiding the trajectories of anyons. While it is easy to understand how to build one-qubit operations, two-qubit operations are more difficult.
In \cite{EntSU2} we suggested an approach to build such operations for a topological quantum computer based on $SU(2)$ Chern-Simons theory with arbitrary level using cabling of knots. In this paper we discuss how this approach should be generalized to the $SU(N)$ case, what the differences are, and which new problems arise.
\end{abstract}

\vspace{.5cm}



\section{Introduction}

Quantum computation has great potential in many different applications if, and it is a big if, we can construct a large and efficient quantum computer. The problem is that as the size of the computer grows, the probabilities of errors inherent in the quantum-computing model also increase, which greatly limits the length of the algorithms that can be used. At the same time, as the size of the computer grows, the size of the algorithms also grows. In this paper we discuss a special approach to quantum computing which tries to minimize errors by construction, namely a topological quantum computer.

The idea behind a topological quantum computer, first suggested by A.~Kitaev \cite{Kit}, is to use topologically protected states. Then the probability of uncontrolled changes of these states is very small. The question then is how to make operations in such a computer. It can be built using anyonic models -- special two-dimensional quasi-particles with non-trivial statistics.  When these particles are exchanged, a non-trivial operation is induced, and we hope that this can be done with little local interaction between the particles by, for example, keeping them at a distance \cite{ReviewTQC}. There are some experimental attempts at building such a computer \cite{ReviewTQC,exp1,exp2,exp3,exp4,exp5,expend}, although there is no substantial success yet.

An effective field-theory description of anyons is given by a three-dimensional topological Chern-Simons gauge theory, and operations corresponding to the braiding of anyon trajectories are quantum $\mathcal{R}$-matrices \cite{KLJ}. Such objects are widely studied in applications to mathematical knot theory, and results from that subject can be applied to the study of a topological quantum computer.

While one-qubit operations for the anyonic topological quantum computer are conceptually simple and discussed in many places \cite{TopoBook,TowTQC,QuantKnots,LargeK,Measure,2qArb}, two-qubit operations are more complex. There are different approaches \cite{Long:2026brj, Long:2025vbb, Levaillant:2015nia, Kaufmann:2023aoc}, but most of them discuss how these operations can be constructed for very specific Chern-Simons level $k$ and rank $N$ of $SU(N)$ group. In \cite{EntSU2} we discussed how these operations can be constructed for the $SU(2)$ anyons with any level $k$. We suggested that we can use cabling \cite{Cable} to entangle anyons from different qubits, and we can specifically choose this braiding pattern so that the fidelity of the corresponding operation would be high, while the operation itself is non-trivial and entangling. In this paper we generalize this approach to the $SU(N)$ case.

\section{Basics of quantum computing}

The idea of quantum computer is based on the fact that quantum systems are (exponentially) hard to model using a classical computer. Therefore if we can make a quantum system with a controlled evolution, it probably can solve some hard problems more effectively than a classical computer. Evolution of a quantum system corresponds to some unitary matrix, therefore the problem which is supposed to be studied on a quantum computer has to be rewritten as a question about unitary matrices.

The universal quantum computer supposedly should be able to solve any such problem (that's why it is called universal). Therefore we should be able to build any unitary matrix (up to some degree of certainty) on such a computer. For this we need to have some basic set of operations from which can construct the required matrix. This set is called a universal set of quantum gates. Basically, for the set to be universal it is enough to have a universal set of one-qubit gates and at least one entangling two-qubit gate. The most common choice of quantum gates is
\begin{equation}
\mathbb{H}=\frac{1}{\sqrt{2}}\left(\begin{array}{cc}1&1\\1&-1\end{array}\right), \ \ \mathbb{T}=\left(\begin{array}{cc}1\\&e^{\frac{i\pi}{4}}\end{array}\right), \ \
CNOT=\left(\begin{array}{cccc}1\\&1\\&&0&1\\&&1&0\end{array}\right).
\end{equation}
This is of course not the only possible choice, and we will consider a different set of operations.

Usually two-qubit operations are much harder for any model of quantum computer, since they imply some kind of interaction between different qubits, which drastically increase the probability of errors. Random errors are a scourge of quantum computer. Due to its nature quantum computer has some probability of errors on a fundamental level, and when we want to evolve controllably it is even harder. Therefore, the idea of topological quantum computer is so luring due to its natural higher resistance to random errors. Ideally in this context we want some model of topological quantum computer with topologically stable qubits and topological operations.

\section{Anyons in Chern-Simons theory}

Anyonic models are effective theories of quasi-particles called anyons which possess non-standard statistics. Namely, exchange of such particles corresponds to an action of some operator, rather than just multiplication by $\pm 1$. These particles supposedly interact with the field of the three-dimensional Chern-Simons theory with some gauge group $\mathcal{G}$
\begin{equation}
S=\frac{k}{4\pi}\text{Tr}\int A\wedge dA+\frac{2}{3}A\wedge A\wedge A.
\end{equation}
We assume that we can somehow control the trajectories of anyons in a topological sense\footnote{It means that we do not care about deviations of the trajectories of the particles, if they do not intersect.}. Then probability amplitude of anyonic processes corresponds to the Wilson-loop average in the Chern-Simons theory:
\begin{equation}
\left<W_V(\mathcal{K})\right>_{CS(\mathcal{G},k)}=\left<\text{Tr}_V\text{Pexp}\oint\limits_{\mathcal{K}}A\right>_{CS(\mathcal{G},k)},
\end{equation}
where $V$ is a representation of the gauge group $\mathcal{G}$ and $\mathcal{K}$ is a contour in three-dimensional space representing the anyon trajectories. Wilson-loop averages are equal to the knot polynomials from the mathematical knot theory \cite{Witt}.

There is a very useful constructive approach to computing these Wilson-loop averages, which involves quantum $\mathcal{R}$ matrices. These matrices are precisely the objects that are supposed to correspond to operations in an anyonic topological quantum computer \cite{KLJ}. There is a large body of work on quantum $\mathcal{R}$-matrices and their application to Wilson-loop averages for the $SU(N)$ group, usually referred to as the modern Reshetikhin-Turaev approach \cite{RT,RT2,RT3,RTus,RTus2}. But the task of applying all of this knowledge to the study of quantum gates for the anyonic quantum computer is still incomplete. In our previous paper \cite{EntSU2} we discussed how we can build the entangling two-qubit gate for the $SU(2)$ anyons for potentially arbitrary values of $k$. In this paper we discuss how to generalize it to the $SU(N)$ case.

It is useful to note that the answers for the Wilson-loop averages are indeed Laurent polynomials in the variables $q$ and $A$, which are related to the parameters of the Chern-Simons theory $k$ and $N$:
\begin{equation}
q=exp\left(\cfrac{2\pi i}{k+N}\right),\ \ \ A=q^N.
\label{eq:Aqdef}
\end{equation}

\section{$\mathcal{R}$-matrices, braids and knot invariants\label{s:Rm}}

The machinery of the Reshetikhin-Turaev approach to Wilson-loop averages and knot polynomials was thoroughly developed in \cite{RT,RT2,RT3,RTus,RTus2,Tabul,DoubleFat,6j-orig,MMS}. Here we will briefly recall its basic ingredients, which we will need in applications to an anyonic quantum computer. According to this approach we should project the contour (or several contours if we have more anyonic particles) onto a two-dimensional plane, which we think of as an $(y,t)$-plane so that one axis represents time. Then each crossing on this projection corresponds to the $\mathcal{R}$-matrix. Also each strand, corresponding to a trajectory of an anyon, carries some representation $V_i$ of the gauge group $\mathcal{G}=SU(N)$. The representations of the group $SU(N)$ can be enumerated by a Young diagram -- non-increasing sequence of natural numbers $V=[v_1\geq v_2\geq v_3\ldots]$. It turns out that, although we start from Chern-Simons theory with a classical Lie gauge group, the answers are expressed in terms of quantities associated with the quantum group $U_q(su(N))$ \cite{Klimyk}, i.e. a deformation of the universal enveloping algebra. Its representations are actually in one-to-one correspondence to the representations of the classical group\footnote{If $q$ is a root-of-unity the situation is actually more complicated, but basically if the degree of the root is large enough (larger than twice the size of the largest representation which appears in the calculations) it does not matter.}. In what follows we will speak simply about representations, keeping in mind the quantum-group interpretation.
\begin{figure}[h!]
\begin{picture}(30,50)(-250,-20)
\put(0,0){\line(1,1){24}}
\put(24,0){\line(-1,1){10}}
\put(0,24){\line(1,-1){10}}
\put(25,10){\hbox{$\mathcal{R}$}}
\put(-5,-10){\hbox{$V_1$}}
\put(19,-10){\hbox{$V_2$}}
\end{picture}
\caption{A graphical description of an $\mathcal{R}$-matrix}
\end{figure}
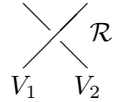
It is important to note that due to the properties of quantum $\mathcal{R}$-matrix its action actually depends on the irreducible representations from the product of two initial representations, and moreover it acts on them by just multiplying them by a corresponding eigenvalue:
\begin{equation}
\begin{array}{l}
V_1\otimes V_2=\sum Q;
\\ \\
\hat{\mathcal{R}} \left(V_1\otimes V_2\right)=\sum \hat{\mathcal{R}} Q=\sum \lambda_Q Q;
\\ \\
\lambda_Q\sim \varepsilon_Q q^{\varkappa_Q},\ \ \ \varkappa_Q=\sum\limits_{(i,j)\in Q} (i-j),
\end{array}
\label{eq:2stR}
\end{equation}
where $\varepsilon_Q$ are the signs of eigenvalues, which depend on whether the representation $Q$ comes from the positive or negative square of representation $V$ \cite{Tabul,HW1,HW2}. Actually the action of the $\mathcal{R}$-matrix is defined up to an overall factor, which will be discussed further in s.\ref{s:fr}, but it is not so important for the discussions in this section.

It is useful to consider $\mathcal{R}$ matrices as operators acting on a braid, i.e. on several parallel strands. On one hand, such an operator is conceptually simple -- it acts on a pair of intersecting strands as was described in (\ref{eq:2stR}), and as a unity operator on all other strands. On the other hand it is useful to modify this statement so that we can easily multiply different $\mathcal{R}$-matrices.

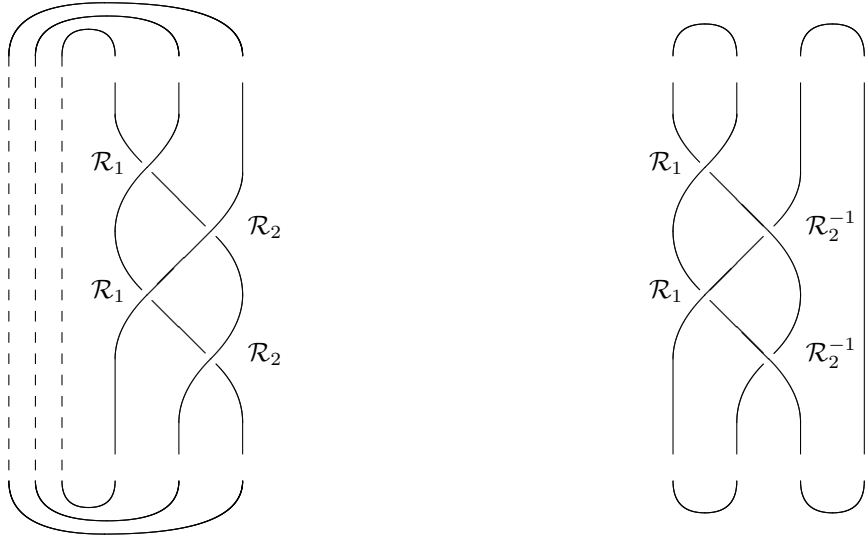
\begin{figure}[h!]
\begin{picture}(30,250)(-120,-200)
\put(0,0){
\put(0,0){\line(0,-1){12}}
\put(24,0){\line(0,-1){12}}
\put(48,0){\line(0,-1){34}}
\qbezier(0,-12)(0,-20)(10,-30)
\qbezier(24,-12)(24,-20)(12,-32)
\qbezier(0,-56)(0,-44)(12,-32)
\put(14,-34){\line(1,-1){20}}
\qbezier(48,-34)(48,-44)(36,-56)
\put(36,-56){\line(-1,-1){24}}
\qbezier(0,-56)(0,-68)(10,-78)
\qbezier(38,-58)(48,-68)(48,-80)
\qbezier(36,-104)(48,-92)(48,-80)
\put(14,-82){\line(1,-1){20}}
\qbezier(12,-80)(0,-92)(0,-104)
\qbezier(36,-104)(24,-116)(24,-128)
\qbezier(38,-106)(48,-116)(48,-128)
\put(0,-104){\line(0,-1){36}}
\put(24,-128){\line(0,-1){12}}
\put(48,-128){\line(0,-1){12}}
%
\put(-9,-33){\hbox{$\mathcal{R}_1$}}
\put(-9,-81){\hbox{$\mathcal{R}_1$}}
\put(50,-57){\hbox{$\mathcal{R}_2$}}
\put(50,-105){\hbox{$\mathcal{R}_2$}}
\multiput(-20,10)(0,-8){20}{\line(0,-1){4}}
\multiput(-30,10)(0,-8){20}{\line(0,-1){4}}
\multiput(-40,10)(0,-8){20}{\line(0,-1){4}}
\qbezier(0,-150)(0,-160)(-10,-160)
\qbezier(-20,-150)(-20,-160)(-10,-160)
\qbezier(24,-150)(24,-165)(-3,-165)
\qbezier(-30,-150)(-30,-165)(-3,-165)
\qbezier(48,-150)(48,-170)(-4,-170)
\qbezier(-40,-150)(-40,-170)(-4,-170)
\qbezier(0,10)(0,20)(-10,20)
\qbezier(-20,10)(-20,20)(-10,20)
\qbezier(24,10)(24,25)(-3,25)
\qbezier(-30,10)(-30,25)(-3,25)
\qbezier(48,10)(48,30)(-4,30)
\qbezier(-40,10)(-40,30)(-4,30)
}
\put(210,0){
\put(0,0){\line(0,-1){12}}
\put(24,0){\line(0,-1){12}}
\put(48,0){\line(0,-1){34}}
\put(72,0){\line(0,-1){140}}
\qbezier(0,-12)(0,-20)(10,-30)
\qbezier(24,-12)(24,-20)(12,-32)
\qbezier(0,-56)(0,-44)(12,-32)
\put(14,-34){\line(1,-1){22}}
\qbezier(48,-34)(48,-44)(38,-54)
\put(34,-58){\line(-1,-1){22}}
\qbezier(0,-56)(0,-68)(10,-78)
\qbezier(36,-56)(48,-68)(48,-80)
\qbezier(38,-102)(48,-92)(48,-80)
\put(14,-82){\line(1,-1){22}}
\qbezier(12,-80)(0,-92)(0,-104)
\qbezier(34,-106)(24,-116)(24,-128)
\qbezier(36,-104)(48,-116)(48,-128)
\put(0,-104){\line(0,-1){36}}
\put(24,-128){\line(0,-1){12}}
\put(48,-128){\line(0,-1){12}}
%
\put(-9,-33){\hbox{$\mathcal{R}_1$}}
\put(-9,-81){\hbox{$\mathcal{R}_1$}}
\put(50,-57){\hbox{$\mathcal{R}_2^{-1}$}}
\put(50,-105){\hbox{$\mathcal{R}_2^{-1}$}}
\qbezier(0,-150)(0,-162)(12,-162)
\qbezier(24,-150)(24,-162)(12,-162)
\qbezier(48,-150)(48,-162)(60,-162)
\qbezier(72,-150)(72,-162)(60,-162)
\qbezier(0,10)(0,22)(12,22)
\qbezier(24,10)(24,22)(12,22)
\qbezier(48,10)(48,22)(60,22)
\qbezier(72,10)(72,22)(60,22)
}
\end{picture}
\caption{Braid description of a trefoil knot with three and four strands. The left braid should be closed as a traditional braid, the right one is a plat representation.\label{f:braids}}
\end{figure}

As a result we get an operator which acts on the whole braid -- its structure is simple: it acts as an $\mathcal{R}$-matrix on the crossing strands and as the identity operator on all other strands. Importantly, the analogue of property (\ref{eq:2stR}) still holds. Let us expand the product of all representations in a braid into irreducible representations:
\begin{equation}
V_1\otimes V_2\otimes V_3\otimes \ldots= \sum M_Q \otimes Q.
\end{equation}
Here $M_Q$ is the multiplicity space of representation $Q$, meaning that the same representation $Q$ can appear several times in such a product. Then all $\mathcal{R}$-matrices can act nontrivially only in the multiplicity space $M_Q$. They do not change type of representation $Q$ and also they act as identity operators inside the representation $Q$ itself.

The diagonal $\mathcal{R}$-matrix has the eigenvalues defined by (\ref{eq:2stR}) where one should look at the product of two crossing representations. However, not all $\mathcal{R}$-matrices commute, therefore not all of them can be made diagonal at the same time. To change from the basis where one $\mathcal{R}$-matrix is diagonal to the basis where another is diagonal one should use change-of-basis matrices, also called Racah matrices in this case. They are related with the associativity property of product of representations. For example we can expand product of three representations in two ways:
\begin{equation}
\begin{array}{cllll}
(V_1\otimes V_2)\otimes V_3 & = & \left(\sum\limits_i Y_i\right)\otimes T_3 & = & \sum Q,
\\ \| &&&& \\
V_1\otimes (V_2\otimes V_3) & = & T_1\otimes \left(\sum\limits_j Y_j\right) & = & \sum Q.
\end{array}
\end{equation}
Then these two expansions are related by the matrix which elements are denoted by:
\begin{equation}
U_{ij}=\left[
\begin{array}{lll}
V_1 & V_2 & Y_i
\\
V_3 & Q & Y_j
\end{array}
\right].
\label{eq:Rdef}
\end{equation}
It is important to note that due to the definition of $q$ from (\ref{eq:Aqdef}) and form of $\mathcal{R}$-matrix eigenvalues, (\ref{eq:2stR}), diagonal $\mathcal{R}$-matrix is unitary. The same is true for the Racah matrix for the large enough $k$ \cite{QuantKnots,LargeK}. It is an orthogonal matrix, therefore it is unitary if it is real, which can be achieved for large enough $k$.

Another important point is how to construct knot polynomials for plat representations (see Fig. \ref{f:braids}). The closure of such a braid, ``hats'' at the bottom and at the top correspond to the projection on the trivial representation $\emptyset$. Therefore two strands connected by a ``hat'' should carry conjugate representations. As it was said before, $\mathcal{R}$-matrices do not change the whole representation in a braid, therefore representation in the whole braid is also trivial. Let us look at the product of two fundamental and two anti-fundamental representations:
\begin{equation}
\begin{array}{cccc}
[1]\otimes\overline{[1]}\otimes [1]\otimes \overline{[1]}=(\emptyset+\text{Adj})\otimes(\emptyset+\text{Adj})=&
\ \ \ 2\cdot\emptyset&+&\ \ \ \ \ \ \ \ \  4\text{Adj}\ \ \ +\ldots
\\
&\nearrow\ \ \ \ \uparrow&&\nearrow\ \ \ \ \ \ \ \uparrow\ \ \ \ \ \ \nwarrow \text{2 times}
\\
&\emptyset\otimes\emptyset\ \ \ \text{Adj}\otimes\text{Adj} && \emptyset\otimes\text{Adj}\ \ \ \text{Adj}\otimes\emptyset\ \ \ \text{Adj}\otimes\text{Adj}
\end{array}
\end{equation}
Thus all the matrices acting in such a braid are of size $2\times 2$. Since $\mathcal{R}$-matrix eigenvalues depend only on the crossing representations, there are two types of them\footnote{Due to the symmetries of the representations theory matrix for the crossing between $\overline{[1]}\times\overline{[1]}$ is also $T$, and for the $\overline{[1]}\times [1]$ is $\overline{T}$} -- the one corresponding to the crossing of $[1]\times [1]$, which we will call $T$, and the one corresponding to the crossing of $[1]\times \overline{[1]}$, which we will call $\overline{T}$. The diagonal form of these matrices is \cite{Tabul,DoubleFat,6j-orig,MMS}:
\begin{equation}
T_{[1]}=\left(\begin{array}{cc}q\xi^{-1} \\ &-(q\xi)^{-1}\end{array}\right),
\ \ \
\overline{T}_{[1]}=\left(\begin{array}{cc}\xi A^{-1} \\ &-\xi\end{array}\right),
\end{equation}
where we use an additional notation
\begin{equation}
\xi=q^{1/N}.
\end{equation}
There are also two Racah matrices, depending on the exact position of the different representations. In this case the Racah matrices are effectively three-strand matrices, since we can make matrices $\mathcal{R}_1$ and $\mathcal{R}_3$ diagonal at the same time. We will denote them as follows, using the notation from (\ref{eq:Rdef}):
\begin{equation}
S=\left[
\begin{array}{lll}
\ [1] & [1] & Y_i
\\
\ \overline{[1]} & [1] & Y_j
\end{array}
\right],\ \ \
\overline{S}=\left[
\begin{array}{lll}
\ [1] & \overline{[1]} & Y_i
\\
\ [1] & [1] & Y_j
\end{array}
\right].
\end{equation}
And they are equal to \cite{Tabul,DoubleFat,6j-orig,MMS}:
\begin{equation}
S_{[1]}=\left(\begin{array}{cc}-\sqrt{\frac{[N-1]_q}{[2]_q[N]_q}} & -\sqrt{\frac{[N+1]_q}{[2]_q[N]_q}} \\ -\sqrt{\frac{[N+1]_q}{[2]_q[N]_q}} &\sqrt{\frac{[N-1]_q}{[2]_q[N]_q}}\end{array}\right),
\ \ \
\overline{S}_{[1]}=\left(\begin{array}{cc} -\frac{1}{[N]_q} & -\frac{\sqrt{[N+1]_q[N-1]_q}}{[N]_q} \\ -\frac{\sqrt{[N+1]_q[N-1]_q}}{[N]_q} & \frac{1}{[N]_q}  \end{array}\right),
\end{equation}
where we use the standard notation for the quantum numbers:
\begin{equation}
[n]_q=\frac{q^n-q^{-n}}{q-q^{-1}}.
\end{equation}
Then the knot polynomial is obtained from a matrix element of a product of $\mathcal{R}$-matrices:
\begin{equation}
\mathcal{B}=\prod \mathcal{R}_i,\ \ \ \mathcal{H}=\frac{q-q^{-1}}{A-A^{-1}}\mathcal{B}_{\emptyset\emptyset}.
\end{equation}

\subsection{Higher symmetric representations\label{s:hrep}}

A similar story holds for higher representations. For symmetric representations defined by a Young diagram $[r]$ matrices will be of the size $(r+1)$ by $(r+1)$. There are general formulae for this case \cite{6j-orig,6j-racah,LargeK}:

\begin{equation}
        \bar S = \epsilon_{\{V_i\}} \sqrt{\dim_q V_{12} \dim_q V_{23}} \cdot
        \left\{\begin{array}{lll}
            V_1 & \bar{V}_2 & V_{12} \\
            V_3 & \bar{V}_4 & V_{23}
        \end{array}\right\},\ \
        S = \epsilon_{\{V_i\}} \sqrt{\dim_{q} V_{12} \dim_{q} V_{23}} \cdot         \left\{\begin{array}{lll}
            V_{1} & V_{2} & V_{12} \\
            \bar{V}_{3} & \bar{V}_{4} & V_{23}
        \end{array}\right\}, \label{rac_2nd}
\end{equation}
where
\begin{equation}
\begin{array}{lcl}
        \dim_q [2n, n^{N-2}] & = & \frac{[N+2n-1]_q [N+n-2]_q!^2}{[n]_q!^2 [N-1]_q! [N-2]_q!},
        \\ \\
        \dim_q [r+n, r-n] & = & \frac{[2n+1]_q [N+r+n-1]_q! [N+r-n-2]_q!}{[N-1]_q! [N-2]_q! [r-n]_q! [r+n+1]_q!},
\end{array}
        \label{q_dim_rrb}
\end{equation}
and
\begin{equation}
\begin{array}{lcl}
        \left\{\begin{array}{lll}
            r & \bar r & i \\
            r & \bar r & j
        \end{array}\right\} & = &
        \frac{[i]_q !^{2}[j]_q !^{2}[r-i]_q ![r-j]_q ![N-1]_q ![N-2]_q !}{[r+i+N-1]_q ![r+j+N-1]_q !} \sum_{z}(-)^{z} \frac{[r+N-1+z]_q !}{[z-i]_q !^{2}[z-j]_q !^{2}[r-z]_q ![i+j-z]_q ![i+j+N-2-z]_q !}
        \\ \\
        \left\{\begin{array}{lll}
            r &  r & i \\
            \bar r & \bar r & j
        \end{array}\right\} & = &
        \frac{[i]_q !^{2}[j]_q !^{2}[r-i]_q ![r-j]_q ![N-1]_q ![N-2]_q !}{[r+i+N-1]_q ![r+j+N-1]_q !} \sum_{z}(-)^{z} \frac{[r+N-1+z]_q !}{[z-i]_q !^{2}[z-j]_q!^{2}[r-z]_q ![i+j-z]_q ![z-j+N-2]_q !}
\end{array}
\end{equation}

In the present paper we will need the formulae for representations $[2]$ and $[1,1]$. The latter can be obtained from the former one using the well-known symmetry of quantum groups, namely that transposition of the Young diagram corresponds to the map $q\rightarrow -q^{-1}$. The formulae for these representations are as follows:
\begin{equation}
\begin{array}{lcl}
T_{[2]}=
\left(\begin{array}{lll}
\cfrac{1}{q^2\xi^4}
\\ &
-\cfrac{1}{\xi^4}
\\ &&
\cfrac{q^4}{\xi^4}
\end{array}\right),
& \ &
T_{[1,1]}=
\left(\begin{array}{lll}
\cfrac{q^2}{\xi^4}
\\ &
-\cfrac{1}{\xi^4}
\\ &&
\cfrac{1}{q^4\xi^4}
\end{array}\right),
\\ \\
\overline{T}_{[2]}=
\left(\begin{array}{lll}
\frac{\xi^4}{q^2A^2}
\\ &
-\frac{\xi^4}{q^2A}
\\ &&
\xi^4
\end{array}\right),
& \ &
\overline{T}_{[1,1]}=
\left(\begin{array}{lll}
\frac{q^2\xi^4}{A^2}
\\ &
-\frac{q^2\xi^4}{A}
\\ &&
\xi^4
\end{array}\right),
\end{array}
\end{equation}
\begin{equation}
S_{[2]}=\left(\begin{array}{ccc}
\sqrt{\frac{[N-1]_q}{[3]_q[N+1]_q}} & \frac{1}{\sqrt{[3]_q}} & \sqrt{\frac{[N+3]_q}{[3]_q[N+1]_q}}
\\ \\
\sqrt{\frac{[N-1]_q[N+2]_q[2]_q}{[N]_q[N+1]_q[4]_q}} & \frac{[2N+2]_q}{[N+1]_q}\sqrt{\frac{[2]_q}{[N]_q[N+2]_q[4]_q}} & -\sqrt{\frac{[N]_q[N+3]_q[2]_q}{[N+1]_q[N+2]_q[4]_q}}
\\ \\
\sqrt{\frac{[N+2]_q[N+3]_q[2]_q}{[N]_q[N+1]_q[3]_q[4]_q}} & -\sqrt{\frac{[N-1]_q[N+3]_q[2]^3_q}{[N]_q[N+2]_q[3]_q[4]_q}} &
\sqrt{\frac{[N-1]_q[N]_q[N+3]_q[2]_q}{[N+1]_q[N+2]_q[N+3]_q[3]_q[4]_q}}
\end{array}\right),
\end{equation}
\begin{equation}
\overline{S}_{[2]}=\left(\begin{array}{ccc}
\frac{[2]_q}{[N]_q[N+1]_q} & \frac{[2]_q}{[N]_q}\sqrt{\frac{[N-1]_q}{[N+1]_q}} & \frac{\sqrt{[N-1]_q[N+3]_q}}{[N+1]_q}
\\ \\
\frac{[2]_q}{[N]_q}\sqrt{\frac{[N-1]_q}{[N+1]_q}} & \frac{[N-1]_q[N+3]_q-1}{[N]_q[N+2]_q} & -\frac{[2]_q}{[N+2]_q}\sqrt{\frac{[N+3]_q}{[N+1]_q}}
\\ \\
\frac{\sqrt{[N-1]_q[N+3]_q}}{[N+1]_q} & -\frac{[2]_q}{[N+2]_q}\sqrt{\frac{[N+3]_q}{[N+1]_q}} &
\frac{[2]_q}{[N+1]_q[N+2]_q}
\end{array}\right),
\end{equation}
\begin{equation}
S_{[1,1]}=\left(\begin{array}{ccc}
\sqrt{\frac{[N+1]_q}{[3]_q[N-1]_q}} & \frac{1}{\sqrt{[3]_q}} & \sqrt{\frac{[N-3]_q}{[3]_q[N-1]_q}}
\\ \\
\sqrt{\frac{[N+1]_q[N-2]_q[2]_q}{[N]_q[N-1]_q[4]_q}} & -\frac{[2N-2]_q}{[N-1]_q}\sqrt{\frac{[2]_q}{[N]_q[N-2]_q[4]_q}} & -\sqrt{\frac{[N]_q[N-3]_q[2]_q}{[N-1]_q[N-2]_q[4]_q}}
\\ \\
\sqrt{\frac{[N-2]_q[N-3]_q[2]_q}{[N]_q[N-1]_q[3]_q[4]_q}} & -\sqrt{\frac{[N+1]_q[N-3]_q[2]^3_q}{[N]_q[N-2]_q[3]_q[4]_q}} &
\sqrt{\frac{[N+1]_q[N]_q[2]_q}{[N-1]_q[N-2]_q[3]_q[4]_q}}
\end{array}\right),
\end{equation}
\begin{equation}
\overline{S}_{[1,1]}=\left(\begin{array}{ccc}
\frac{[2]_q}{[N]_q[N-1]_q} & -\frac{[2]_q}{[N]_q}\sqrt{\frac{[N+1]_q}{[N-1]_q}} & -\frac{\sqrt{[N+1]_q[N-3]_q}}{[N-1]_q}
\\ \\
-\frac{[2]_q}{[N]_q}\sqrt{\frac{[N+1]_q}{[N-1]_q}} & \frac{[N+1]_q[N-3]_q-1}{[N]_q[N-2]_q} & -\frac{[2]_q}{[N-2]_q}\sqrt{\frac{[N-3]_q}{[N-1]_q}}
\\ \\
-\frac{\sqrt{[N+1]_q[N-3]_q}}{[N-1]_q} & -\frac{[2]_q}{[N-2]_q}\sqrt{\frac{[N-3]_q}{[N-1]_q}} &
\frac{[2]_q}{[N-1]_q[N-2]_q}
\end{array}\right),
\end{equation}

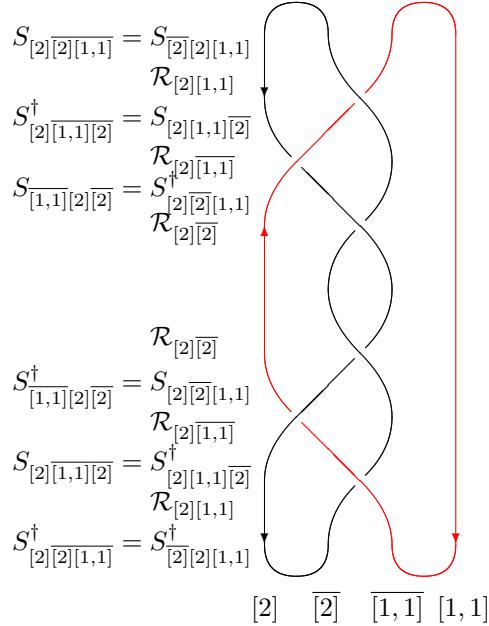
\begin{figure}[h!]
\begin{picture}(30,230)(-120,-160)
\put(130,0){
\put(0,64){\vector(0,-1){24}}
\qbezier(10,18)(0,28)(0,40)
\qbezier(36,40)(24,52)(24,64)
\qbezier(48,16)(48,28)(36,40)
\put(36,-8){\line(-1,1){22}}
\qbezier(38,-6)(48,4)(48,16)
\qbezier(36,-8)(48,-20)(48,-32)
\qbezier(34,-10)(24,-20)(24,-32)
\qbezier(36,-56)(24,-44)(24,-32)
\qbezier(48,-32)(48,-44)(38,-54)
\put(34,-58){\line(-1,-1){22}}
\qbezier(36,-56)(48,-68)(48,-80)
\qbezier(38,-102)(48,-92)(48,-80)
\qbezier(12,-80)(0,-92)(0,-104)
\qbezier(34,-106)(24,-116)(24,-128)
%
%
\put(0,-104){\vector(0,-1){24}}
\qbezier(0,-128)(0,-140)(12,-140)
\qbezier(24,-128)(24,-140)(12,-140)
{\color{red}
\qbezier(48,64)(48,76)(60,76)
\qbezier(72,64)(72,76)(60,76)
\qbezier(38,42)(48,52)(48,64)
\put(12,16){\line(1,1){22}}
\qbezier(12,16)(0,4)(0,-8)
\put(0,-56){\vector(0,1){48}}
\put(72,64){\vector(0,-1){192}}
\qbezier(0,-56)(0,-68)(10,-78)
\put(14,-82){\line(1,-1){22}}
\qbezier(36,-104)(48,-116)(48,-128)
\qbezier(48,-128)(48,-140)(60,-140)
\qbezier(72,-128)(72,-140)(60,-140)
}
\qbezier(0,64)(0,76)(12,76)
\qbezier(24,64)(24,76)(12,76)
\put(-95,60){\hbox{$S_{[2]\overline{[2]}\overline{[1,1]}}=S_{\overline{[2]}{[2]}{[1,1]}}$}}
\put(-43,45){\hbox{$\mathcal{R}_{[2]{[1,1]}}$}}
\put(-95,30){\hbox{$S^{\dagger}_{[2]\overline{[1,1]}\overline{[2]}}=S_{{[2]}{[1,1]}\overline{[2]}}$}}
\put(-43,15){\hbox{$\mathcal{R}_{[2]\overline{[1,1]}}$}}
\put(-95,3){\hbox{$S_{\overline{[1,1]}[2]\overline{[2]}}=S^{\dagger}_{[2]\overline{[2]}{[1,1]}}$}}
\put(-43,-10){\hbox{$\mathcal{R}_{[2]\overline{[2]}}$}}
\put(-43,-53){\hbox{$\mathcal{R}_{[2]\overline{[2]}}$}}
\put(-95,-70){\hbox{$S^{\dagger}_{\overline{[1,1]}[2]\overline{[2]}}=S_{[2]\overline{[2]}{[1,1]}}$}}
\put(-43,-85){\hbox{$\mathcal{R}_{[2]\overline{[1,1]}}$}}
\put(-95,-100){\hbox{$S_{[2]\overline{[1,1]}\overline{[2]}}=S^{\dagger}_{{[2]}{[1,1]}\overline{[2]}}$}}
\put(-43,-115){\hbox{$\mathcal{R}_{[2][1,1]}$}}
\put(-95,-130){\hbox{$S^{\dagger}_{[2]\overline{[2]}\overline{[1,1]}}=S^{\dagger}_{\overline{[2]}{[2]}{[1,1]}}$}}
\put(-5,-155){\hbox{$[2]$}}
\put(18,-155){\hbox{$\overline{[2]}$}}
\put(40,-155){\hbox{$\overline{[1,1]}$}}
\put(65,-155){\hbox{$[1,1]$}}
}
\end{picture}
\caption{Plat representation of a link L6a1. On the left the corresponding $\mathcal{R}$ and Racah matrices are written in the order in which they are multiplied from top to bottom. In the provided example the black component is colored with the symmetric representation $[2]$ and the red component with representation $[1,1]$.
\label{f:2brLink}}
\end{figure}

Finally, the last configuration we will need is a plat representation of a link (see Fig.\ref{f:2brLink}). In this case there are two different components in the plat diagram of the link, which can carry different representations. If they carry the same representations, then the matrices are the same as discussed above. However, if they are different, the matrices are more complicated. In the case we are interested in the expansion of the product of representations is as follows:
\begin{equation}
[2]\otimes[1,1]\otimes \overline{[2]}\otimes\overline{[1,1]}=([3,1]+[2,1,1])\otimes(\overline{[3,1]}+\overline{[2,1,1]})=2\cdot \emptyset+\ldots.
\end{equation}

Therefore all the matrices are again of the size 2 by 2. $\mathcal{R}$-matrices depend on the representations on the crossing strands, and since we have $4$ different representations, there are $4$ different $\mathcal{R}$-matrices\footnote{There are also two types of symmetries which takes care of the other 8 possible combinations of representations. The first one is that if we conjugate all the representations the matrix would be the same, e.g. $\mathcal{R}_{[2][1,1]}=\mathcal{R}_{\overline{[2][1,1]}}$. The second one is that if we permute two representaions, matrix will also be the same, e.g. $\mathcal{R}_{[2][1,1]}=\mathcal{R}_{[1,1][2]}$.}. The diagonal form of all these matrices can be find using (\ref{eq:2stR}) or (\ref{eq:vfr}):
\begin{equation}
\begin{array}{lcl}
\mathcal{R}_{[2]\overline{[2]}}=
\left(\begin{array}{ll}
\cfrac{\xi^4}{A^2q^2}
\\ &
-\cfrac{\xi^4}{Aq^2}
\end{array}\right),
&\ &
\mathcal{R}_{[1,1]\overline{[1,1]}}=
\left(\begin{array}{ll}
\cfrac{\xi^4q^2}{A^2}
\\ &
-\cfrac{\xi^4q^2}{A}
\end{array}\right),
\\
\mathcal{R}_{[2][1,1]}=
\left(\begin{array}{ll}
\cfrac{1}{q^2\xi^4}
\\ &
\cfrac{q^2}{\xi^4}
\end{array}\right),
&\ &
\mathcal{R}_{[2]\overline{[1,1]}}=
\left(\begin{array}{ll}
\cfrac{\xi^4}{A}
\\ &
\xi^4
\end{array}\right)
\end{array}
\end{equation}
As was already said all Racah matrices are essentially three-strand matrices. They also depend on the placement of representations in the braid, just like the $\mathcal{R}$-matrices did. We can use formulae from section 10.3.2 in \cite{Cable} to find them\footnote{We list only Racah matrices which are sufficiently different, employing their symmetries:
$$S_{V_1V_2V_3}=S_{\overline{V_1}\overline{V_2}\overline{V_3}}=S^{\dagger}_{V_3V_2V_1}=S^{\dagger}_{V_2V_3V_4}=S_{V^T_1V^T_2V^T_3},$$
where $V_4$ is the fourth representation in a plat braid.}:
\begin{equation}
\begin{array}{c}
S_{\overline{[2]}[2][1,1]}=\left(\begin{array}{lcl}
-\sqrt{\frac{[2]_q[N-2]_q}{[4]_q[N]_q}} && \sqrt{\frac{[2]_q[N+2]_q}{[4]_q[N]_q}}
\\ \\
\sqrt{\frac{[2]_q[N+2]_q}{[4]_q[N]_q}} && \sqrt{\frac{[2]_q[N-2]_q}{[4]_q[N]_q}}
\end{array}\right),\ \ \
S_{{[2]}\overline{[2]}[1,1]}=\left(\begin{array}{ccc}
-{\frac{\sqrt{[N+2]_q[N-2]_q}}{[N]_q}} && {\frac{[2]_q}{[N]_q}}
\\ \\
{\frac{[2]_q}{[N]_q}} && \frac{\sqrt{[N+2]_q[N-2]_q}}{[N]_q}
\end{array}\right)
\\ \\
S_{{[2]}[1,1]\overline{[2]}}=\left(\begin{array}{lcl}
-\sqrt{\frac{[2]_q[N+2]_q}{[4]_q[N]_q}} && \sqrt{\frac{[2]_q[N-2]_q}{[4]_q[N]_q}}
\\ \\
\sqrt{\frac{[2]_q[N-2]_q}{[4]_q[N]_q}} && \sqrt{\frac{[2]_q[N+2]_q}{[4]_q[N]_q}}
\end{array}\right).
\end{array}
\end{equation}
Let us recall an important fact about all the listed matrices. If we fix
\begin{equation}
q=\text{exp}\left(\frac{2\pi i}{k+N}\right),\ \ A=q^N,\ \ \xi=q^{1/N},
\end{equation}
as is required by the Chern-Simons theory, all of them are unitary if $k$ is large enough (namely if all the entries in the Racah matrices are real).

\subsection{Adjoint representation\label{s:adj}}

Another case important for the our discussion is adjoint representation. Adjoint representation appears in the product of fundamenatl and antifundamental representation. It is self-conjugate therefore all the strands even in the 4-plat case carry adjoint representations. Matrices for the adjoint representations are thoroughly discussed \cite{UniRac} not only for the $U_q(sl_N)$ quantum group but also for general Vogel-like algebra.

An expansion into irreducible representations can be written as follows:
\begin{equation}
\text{Adj}\otimes\text{Adj}=\emptyset\oplus Y_2\oplus Y_2^{\prime}\oplus Y_2^{\prime\prime}\oplus \text{Adj}\oplus X_2,
\end{equation}
where for the $U_q(sl_N)$ case these representations are equal to
\begin{equation}
Y_2=[4,2^{N-2}],\ \ Y_2^{\prime}=[2,2,1^{N-4}],\ \ Y_2^{\prime\prime}=\text{Adj},\ \ X_2=[3,1^{N-3}]\oplus[3,3,2^{N-3}],
\end{equation}
where $n^N$ denotes $n$ repeated $N$ times, e.g. $[2,1^3]=[2,1,1,1]$.
Corresponding diagonal $\mathcal{R}$-matrix is equal to
\begin{equation}
T=\text{diag}\left[A^{-2},q^2,q^{-2},A^{-1},-A^{-1},-1\right].
\end{equation}
Since adjoint representation is self-conjugate, there is only one $\mathcal{R}$-matrix in the $4$-plat structure. Also there is only one Racah matrix $\mathfrak{S}=S_{\text{Adj}}$ found in \cite{UniRac}. It is symmetric and its elements are equal to
\begin{equation}
\begin{array}{lll}
\mathfrak{S}_{11}=\cfrac{1}{[N-1]_q[N+1]_q},
&
\mathfrak{S}_{22}=\cfrac{[N+1]_q[N]_q+[2]_q}{[2]_q^2[N+1]_q[N+2]_q},
&
\mathfrak{S}_{33}=\cfrac{[N]_q[N-1]_q+[2]_q}{[2]_q^2[N-1]_q[N-2]_q},
\\ \\
\mathfrak{S}_{44}=\cfrac{A^2 q^2 - 3 A q^4 + 4 A q^2 - 3 A + q^2}{2(A-q^{-2})(A-q^2)},
&
\mathfrak{S}_{55}=\cfrac{A^2 q^2 + 3 A q^4 - 4 A q^2 + 3 A + q^2}{2(A+q^2)(Aq^2+1)},
&
\mathfrak{S}_{66}=\cfrac{2}{[2]_q^2},
\\ \\
\mathfrak{S}_{12}=-\cfrac{[N]_q}{[2]_q[N+1]_q}\sqrt{\cfrac{[N+3]_q}{[N-1]_q}},
&
\mathfrak{S}_{13}=-\cfrac{[N]_q}{[2]_q[N-1]_q}\sqrt{\cfrac{[N-3]_q}{[N+1]_q}},
\\ \\
\mathfrak{S}_{14}=\mathfrak{S}_{15}=-\cfrac{1}{\sqrt{[N+1]_q[N-1]_q}},
&
\mathfrak{S}_{16}=\cfrac{1}{[2]_q}\sqrt{\cfrac{2[N-2]_q[N+2]_q}{[N-1]_q[N+1]_q}},
&
\mathfrak{S}_{23}=\cfrac{1}{[2]^2_q}\sqrt{\cfrac{[N-3]_q[N+3]_q}{[N-1]_q[N+1]_q}},
\\ \\
\mathfrak{S}_{24}=\cfrac{q(A-1)}{(Aq^2-1)[2]_q}\sqrt{\cfrac{[N+3]_q}{[N+1]_q}},
&
\mathfrak{S}_{25}=-\cfrac{q(A+1)}{(Aq^2+1)[2]_q}\sqrt{\cfrac{[N+3]_q}{[N+1]_q}},
&
\mathfrak{S}_{26}=\cfrac{1}{[2]^2_q}\sqrt{\cfrac{2[N-2]_q[N+3]_q}{[N+1]_q[N+2]_q}},
\\ \\
\mathfrak{S}_{34}=-\cfrac{q(A-1)}{(A-q^2)[2]_q}\sqrt{\cfrac{[N-3]_q}{[N-1]_q}},
&
\mathfrak{S}_{35}=\cfrac{q(A+1)}{(A+q^2)[2]_q}\sqrt{\cfrac{[N-3]_q}{[N-1]_q}},
&
\mathfrak{S}_{36}=\cfrac{1}{[2]^2_q}\sqrt{\cfrac{2[N-3]_q[N+2]_q}{[N-1]_q[N-2]_q}},
\\ \\
\mathfrak{S}_{45}=\cfrac{1}{2},
&
\mathfrak{S}_{46}=-\cfrac{1}{[2]_q}\sqrt{\cfrac{(A+q^2)(Aq^2+1)}{2(A-q^2)(Aq^2-1)}},
&
\mathfrak{S}_{56}=\cfrac{1}{[2]_q}\sqrt{\cfrac{(A-q^2)(Aq^2-1)}{2(A+q^2)(Aq^2+1)}},
\end{array}
\end{equation}
These matrices are also unitary under the appropriate choice of parameters.

\subsection{Cabling of knots}

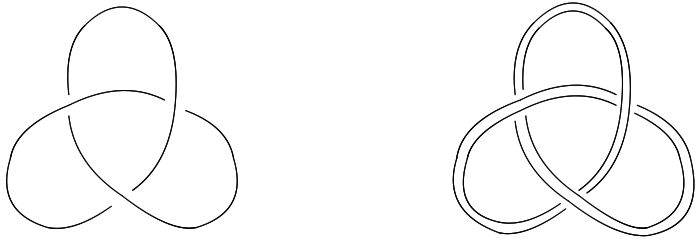
\begin{figure}[h!]
\begin{picture}(250,110)(-55,-55)
\put(80,10){
\qbezier(-20,0)(0,10)(16,2)
\qbezier(-20,-4)(-18,-22)(0,-34)
\qbezier(20,0)(18,-22)(4,-32)
\qbezier(20,0)(22,22)(14,30)
\qbezier(-20,4)(-22,22)(-14,30)
\qbezier(-14,30)(0,44)(14,30)
\qbezier(28,-46)(48,-40)(42,-20)
\qbezier(-28,-46)(-48,-40)(-42,-20)
\qbezier(24,-2)(40,-8)(42,-20)
\qbezier(-20,0)(-40,-8)(-42,-20)
\qbezier(0,-34)(18,-48)(28,-46)
\qbezier(-4,-38)(-18,-48)(-28,-46)
}
\put(250,10){
\qbezier(-20,2)(0,12)(16,4)
\qbezier(-20,-2)(0,8)(16,0)
\qbezier(-18,-4)(-16,-22)(2,-34)
\qbezier(-22,-6)(-20,-24)(-2,-36)
\qbezier(18,0)(16,-22)(2,-32)
\qbezier(21,-1)(19,-23)(5,-33)
\qbezier(18,0)(20,22)(12,29)
\qbezier(21,-1)(23,21)(14,31)
\qbezier(-19,6)(-20,22)(-13,29)
\qbezier(-22,4)(-24,22)(-15,31)
\qbezier(-13,29)(0,42)(12,29)
\qbezier(-15,31)(0,46)(14,31)
\qbezier(28,-46)(46,-40)(40,-20)
\qbezier(28,-49)(50,-45)(44,-20)
\qbezier(-28,-48)(-50,-40)(-44,-20)
\qbezier(-28,-44)(-46,-40)(-40,-20)
\qbezier(24,0)(41,-7)(44,-20)
\qbezier(23,-3)(38,-10)(40,-20)
\qbezier(-20,2)(-42,-6)(-44,-20)
\qbezier(-20,-2)(-38,-10)(-40,-20)
\qbezier(2,-34)(20,-48)(28,-46)
\qbezier(-2,-36)(16,-50)(28,-49)
\qbezier(-3,-39)(-18,-50)(-28,-48)
\qbezier(-5,-37)(-18,-46)(-28,-44)
}
\end{picture}
\caption{A trefoil knot on the left and a $2$-cabled trefoil knot on the right. The strand is replaced with two parallel strands.}
\label{f:Cable}
\end{figure}

The discussed approach to knot theory is heavily based on the representation theory of quantum groups. And it appears that knot polynomials behave much like the representations of quantum groups and their dimensions. Representations can be multiplied and the product of them expanded into product of other irreducible representations\footnote{Some representations $Q$ can appear several times, the multiplicity space does not play a role in the discussion of this subsection}:
\begin{equation}
V_1\otimes V_2\otimes V_3\otimes \ldots= \sum Q.
\end{equation}

The idea of cabling \cite{Cable} is that if we take a knot made of cable, several parallel strands, see Fig.\ref{f:Cable} then the resulting polynomial is a sum of the polynomials in higher representations:
\begin{equation}
\mathcal{H}_{V_1\otimes V_2\otimes V_3\otimes \ldots}= \sum \mathcal{H}_Q.
\end{equation}

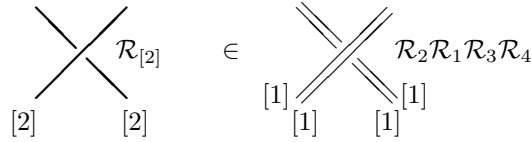
\begin{figure}[h!]
\begin{picture}(100,60)(-180,-20)
\put(0,0){
\thicklines{\put(-5,-5){\line(1,1){34}}
\put(29,-5){\line(-1,1){15}}
\put(-5,29){\line(1,-1){15}}
}
\put(25,10){\hbox{$\mathcal{R}_{[2]}$}}
\put(-15,-17){\hbox{$[2]$}}
\put(27,-17){\hbox{$[2]$}}
}
\put(65,10){\hbox{$\in$}}
\put(100,0){
\put(-7,-5){\line(1,1){34}}
\put(-5,-7){\line(1,1){34}}
\put(29,-7){\line(-1,1){15}}
\put(31,-5){\line(-1,1){15}}
\put(-7,29){\line(1,-1){15}}
\put(-5,31){\line(1,-1){15}}
\put(30,10){\hbox{$\mathcal{R}_{2}\mathcal{R}_{1}\mathcal{R}_{3}\mathcal{R}_{4}$}}
\put(-9,-17){\hbox{$[1]$}}
\put(22,-17){\hbox{$[1]$}}
\put(-20,-7){\hbox{$[1]$}}
\put(32,-7){\hbox{$[1]$}}
}
\end{picture}
\caption{Cabled $\mathcal{R}$-matrix. The $\mathcal{R}$-matrix in representation $[2]$ can be built from a product of four $\mathcal{R}$-matrices in the fundamental representation on the right. Indices of the matrices on the right indicate which matrix in the four-strand braid should be taken.}
\label{f:CR}
\end{figure}

This relation in fact also holds at the level of $\mathcal{R}$-matrices -- matrices in higher representations can be expressed through the matrices in lower representations. For example, see Fig.\ref{f:CR}, if we take a crossing of two strand cables, then the matrices in representations $[2]$ and $[1,1]$ can be made from the product of four $\mathcal{R}$-matrices in fundamental representation after inserting the corresponding projectors in the cables.

Our approach to a topological quantum computer relies heavily on the cabling procedure, as we will discuss below.

\section{One-qubit topological operations}


The idea of a topological quantum computer is based on using the evolution of anyons as a medium for quantum calculations. Due to the topological nature of the states and the non-local character of the interaction (interaction by intertwining of trajectories), such a computer should be topologically stable and have a low probability of random errors.

Operations in such a computer are $\mathcal{R}$-matrices (more precisely, diagonal $\mathcal{R}$-matrices and Racah matrices). They are unitary and therefore are valid as quantum computer operators. The smallest number of anyons for which there are at least two non-trivial operations is three. For one anyon there are no operations at all and for two anyons there is only one $\mathcal{R}$-matrix, which therefore cannot provide a universal set of one-qubit gates. Three anyons are not a convenient number of particles, since it is easier to create pairs of particles. Therefore it is logical to define one-qubit operation as a four strand braid with a plat closure. The matrices acting in such a braid were defined in section \ref{s:Rm}. In fact, as we can see from those formulae, the matrices for representation $[r]$ are of size $(r+1)\times (r+1)$ and, therefore, can be used as one-qudit operations. In \cite{LargeK} it was proven that they indeed provide a universal set of one-qudit operations if $k$ is large enough.

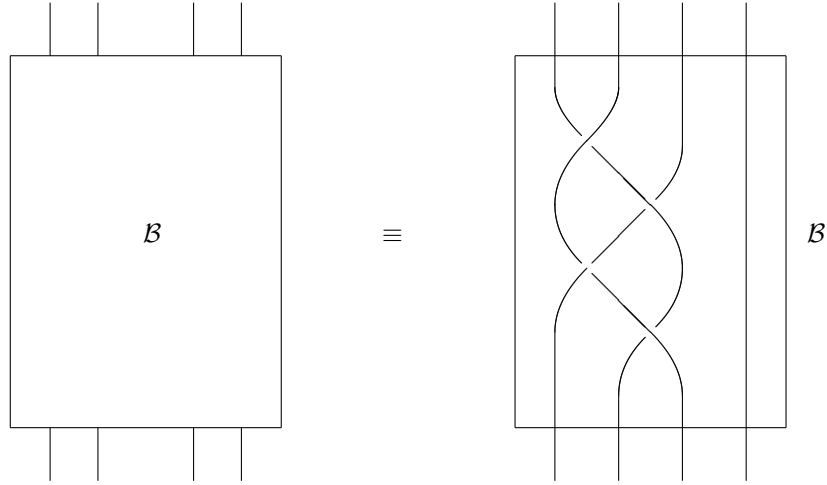
\begin{figure}[h!]
\begin{picture}(30,180)(-120,-160)
\put(0,0){
%
\put(35,-70){\hbox{$\mathcal{B}$}}
\put(0,-140){\line(0,-1){20}}
\put(18,-140){\line(0,-1){20}}
\put(54,-140){\line(0,-1){20}}
\put(72,-140){\line(0,-1){20}}
\put(0,0){\line(0,1){20}}
\put(18,0){\line(0,1){20}}
\put(54,0){\line(0,1){20}}
\put(72,0){\line(0,1){20}}
\put(-15,0){\line(1,0){102}}
\put(-15,0){\line(0,-1){140}}
\put(87,0){\line(0,-1){140}}
\put(-15,-140){\line(1,0){102}}
}
\put(125,-70){\hbox{$\equiv$}}
\put(190,0){
\put(0,0){\line(0,-1){12}}
\put(24,0){\line(0,-1){12}}
\put(48,0){\line(0,-1){34}}
\put(72,0){\line(0,-1){140}}
\qbezier(0,-12)(0,-20)(10,-30)
\qbezier(24,-12)(24,-20)(12,-32)
\qbezier(0,-56)(0,-44)(12,-32)
\put(14,-34){\line(1,-1){22}}
\qbezier(48,-34)(48,-44)(38,-54)
\put(34,-58){\line(-1,-1){20}}
\qbezier(0,-56)(0,-68)(10,-78)
\qbezier(36,-56)(48,-68)(48,-80)
\qbezier(38,-102)(48,-92)(48,-80)
\put(14,-82){\line(1,-1){22}}
\qbezier(12,-80)(0,-92)(0,-104)
\qbezier(34,-106)(24,-116)(24,-128)
\qbezier(36,-104)(48,-116)(48,-128)
\put(0,-104){\line(0,-1){36}}
\put(24,-128){\line(0,-1){12}}
\put(48,-128){\line(0,-1){12}}
%
\put(95,-70){\hbox{$\mathcal{B}$}}
\put(0,-140){\line(0,-1){20}}
\put(24,-140){\line(0,-1){20}}
\put(48,-140){\line(0,-1){20}}
\put(72,-140){\line(0,-1){20}}
\put(0,0){\line(0,1){20}}
\put(24,0){\line(0,1){20}}
\put(48,0){\line(0,1){20}}
\put(72,0){\line(0,1){20}}
\put(-15,0){\line(1,0){102}}
\put(-15,0){\line(0,-1){140}}
\put(87,0){\line(0,-1){140}}
\put(-15,-140){\line(1,0){102}}
}
\end{picture}
\caption{In a topological quantum computer, a one-qubit operation corresponds to a $4$-strand braid. From the braid we can construct a matrix $\mathcal{B}$ as a product of $\mathcal{R}$-matrices in the braid. A one-qubit operation is therefore described by this matrix, and we will denote it as a box with $\mathcal{B}$ inside. Braid on the picture is of course just an example, operations are defined for any braid $\mathcal{B}$, and the universality holds if we consider all of them.\label{f:1qb}}
\end{figure}

We will draw one qubit operations as on Fig.\ref{f:1qb}.

\section{Two-qubit operations}

\begin{figure}[h!]
\begin{subfigure}{.3\textwidth}
\begin{picture}(150,190)(-30,-75)
\put(0,-50){
\put(0,0){\line(1,0){25}}
\put(0,0){\line(0,1){35}}
\put(25,35){\line(-1,0){25}}
\put(25,35){\line(0,-1){35}}
\put(5,13){\hbox{$\mathcal{B}^{(1)}_{1}$}}
\qbezier(3,0)(3,-8)(6,-8)
\qbezier(9,0)(9,-8)(6,-8)
\qbezier(16,0)(16,-8)(19,-8)
\qbezier(22,0)(22,-8)(19,-8)
}
\put(0,76){
\put(0,0){\line(1,0){25}}
\put(0,0){\line(0,1){35}}
\put(25,35){\line(-1,0){25}}
\put(25,35){\line(0,-1){35}}
\put(5,13){\hbox{$\mathcal{B}^{(1)}_{n}$}}
\qbezier(3,35)(3,43)(6,43)
\qbezier(9,35)(9,43)(6,43)
\qbezier(16,35)(16,43)(19,43)
\qbezier(22,35)(22,43)(19,43)
}
\put(36,-50){
\put(0,0){\line(1,0){25}}
\put(0,0){\line(0,1){35}}
\put(25,35){\line(-1,0){25}}
\put(25,35){\line(0,-1){35}}
\put(5,13){\hbox{$\mathcal{B}^{(2)}_1$}}
\qbezier(3,0)(3,-8)(6,-8)
\qbezier(9,0)(9,-8)(6,-8)
\qbezier(16,0)(16,-8)(19,-8)
\qbezier(22,0)(22,-8)(19,-8)
}
\put(36,76){
\put(0,0){\line(1,0){25}}
\put(0,0){\line(0,1){35}}
\put(25,35){\line(-1,0){25}}
\put(25,35){\line(0,-1){35}}
\put(5,13){\hbox{$\mathcal{B}^{(2)}_n$}}
\qbezier(3,35)(3,43)(6,43)
\qbezier(9,35)(9,43)(6,43)
\qbezier(16,35)(16,43)(19,43)
\qbezier(22,35)(22,43)(19,43)
}

\put(9,-15){\line(0,1){28}}
\put(3,-15){\line(0,1){28}}
\put(57,-15){\line(0,1){28}}
\put(51,-15){\line(0,1){28}}
\put(16,-15){\line(0,1){28}}
\put(22,-15){\line(0,1){28}}
\put(39,-15){\line(0,1){28}}
\put(45,-15){\line(0,1){28}}
\put(9,48){\line(0,1){28}}
\put(3,48){\line(0,1){28}}
\put(57,48){\line(0,1){28}}
\put(51,48){\line(0,1){28}}
\put(16,48){\line(0,1){28}}
\put(22,48){\line(0,1){28}}
\put(39,48){\line(0,1){28}}
\put(45,48){\line(0,1){28}}
\put(0,13){
\put(0,0){\line(1,0){25}}
\put(0,0){\line(0,1){7}}
\put(25,0){\line(0,1){7}}
\put(25,35){\line(-1,0){25}}
\put(25,35){\line(0,-1){7}}
\put(0,35){\line(0,-1){7}}
\put(5,9){\hbox{$\,\cdot\cdot\cdot$}}
\put(5,20){\hbox{$\,\cdot\cdot\cdot$}}
}
\put(36,13){
\put(0,0){\line(1,0){25}}
\put(0,0){\line(0,1){7}}
\put(25,0){\line(0,1){7}}
\put(25,35){\line(-1,0){25}}
\put(25,35){\line(0,-1){7}}
\put(0,35){\line(0,-1){7}}
\put(5,9){\hbox{$\,\cdot\cdot\cdot$}}
\put(5,20){\hbox{$\,\cdot\cdot\cdot$}}
}
\end{picture}
\caption{Two unentangled qubits
\label{2free}}
\end{subfigure}%
 \begin{subfigure}{.35\textwidth}
\begin{picture}(200,190)(-50,-75)
\put(0,-50){
\put(0,0){\line(1,0){25}}
\put(0,0){\line(0,1){35}}
\put(25,35){\line(-1,0){25}}
\put(25,35){\line(0,-1){35}}
\put(5,13){\hbox{$\mathcal{B}^{(1)}_{1}$}}
\qbezier(3,0)(3,-8)(6,-8)
\qbezier(9,0)(9,-8)(6,-8)
\qbezier(16,0)(16,-8)(19,-8)
\qbezier(22,0)(22,-8)(19,-8)
}
\put(0,75){
\put(0,0){\line(1,0){25}}
\put(0,0){\line(0,1){35}}
\put(25,35){\line(-1,0){25}}
\put(25,35){\line(0,-1){35}}
\put(5,13){\hbox{$\mathcal{B}^{(1)}_{2}$}}
\qbezier(3,35)(3,43)(6,43)
\qbezier(9,35)(9,43)(6,43)
\qbezier(16,35)(16,43)(19,43)
\qbezier(22,35)(22,43)(19,43)
}
\put(32,-50){
\put(0,0){\line(1,0){25}}
\put(0,0){\line(0,1){35}}
\put(25,35){\line(-1,0){25}}
\put(25,35){\line(0,-1){35}}
\put(5,13){\hbox{$\mathcal{B}^{(2)}_1$}}
\qbezier(3,0)(3,-8)(6,-8)
\qbezier(9,0)(9,-8)(6,-8)
\qbezier(16,0)(16,-8)(19,-8)
\qbezier(22,0)(22,-8)(19,-8)
}
\put(32,75){
\put(0,0){\line(1,0){25}}
\put(0,0){\line(0,1){35}}
\put(25,35){\line(-1,0){25}}
\put(25,35){\line(0,-1){35}}
\put(5,13){\hbox{$\mathcal{B}^{(2)}_2$}}
\qbezier(3,35)(3,43)(6,43)
\qbezier(9,35)(9,43)(6,43)
\qbezier(16,35)(16,43)(19,43)
\qbezier(22,35)(22,43)(19,43)
}

\put(1.5,-15){\line(0,1){5}}
\put(7.5,-15){\line(0,1){5}}

\put(1.5,-10){\qbezier(0,0)(0,2.5)(3,5)\qbezier(3,5)(6,7.5)(6,10)\qbezier(6,0)(6,2.5)(4,4)\qbezier(2,6)(0,7,5)(0,10)}
\put(1.5,0){\qbezier(0,0)(0,2.5)(3,5)\qbezier(3,5)(6,7.5)(6,10)\qbezier(6,0)(6,2.5)(4,4)\qbezier(2,6)(0,7,5)(0,10)}

\put(1.5,10){\line(0,1){5}}
\put(7.5,10){\line(0,1){5}}

\put(1.5,45){\qbezier(0,0)(0,2.5)(3,5)\qbezier(3,5)(6,7.5)(6,10)\qbezier(6,0)(6,2.5)(4,4)\qbezier(2,6)(0,7,5)(0,10)}
\put(1.5,55){\qbezier(0,0)(0,2.5)(3,5)\qbezier(3,5)(6,7.5)(6,10)\qbezier(6,0)(6,2.5)(4,4)\qbezier(2,6)(0,7,5)(0,10)}
\put(1.5,65){\qbezier(0,0)(0,2.5)(3,5)\qbezier(3,5)(6,7.5)(6,10)\qbezier(6,0)(6,2.5)(4,4)\qbezier(2,6)(0,7,5)(0,10)}
\put(55.5,-15){\line(0,1){10}}
\put(49.5,-15){\line(0,1){10}}
\put(55.5,5){\line(0,1){40}}
\put(49.5,5){\line(0,1){10}}
\put(39.5,15){\qbezier(0,0)(0,4)(5,7)\qbezier(5,7)(10,10)(10,14)\qbezier(10,0)(10,4)(6,6)\qbezier(4,8)(0,10)(0,14)}
\put(39.5,29){\qbezier(0,0)(0,4)(5,7)\qbezier(5,7)(10,10)(10,14)\qbezier(10,0)(10,4)(6,6)\qbezier(4,8)(0,10)(0,14)}
\put(49.5,43){\line(0,1){2}}
\put(49.5,-5){\qbezier(0,0)(0,2.5)(3,5)\qbezier(3,5)(6,7.5)(6,10)\qbezier(6,0)(6,2.5)(4,4)\qbezier(2,6)(0,7,5)(0,10)}
\put(49.5,45){\qbezier(0,0)(0,2.5)(3,5)\qbezier(3,5)(6,7.5)(6,10)\qbezier(6,0)(6,2.5)(4,4)\qbezier(2,6)(0,7,5)(0,10)}
\put(49.5,55){\qbezier(0,0)(0,2.5)(3,5)\qbezier(3,5)(6,7.5)(6,10)\qbezier(6,0)(6,2.5)(4,4)\qbezier(2,6)(0,7,5)(0,10)}
\put(49.5,65){\qbezier(0,0)(0,2.5)(3,5)\qbezier(3,5)(6,7.5)(6,10)\qbezier(6,0)(6,2.5)(4,4)\qbezier(2,6)(0,7,5)(0,10)}
\put(33.5,15){\line(0,1){30}}\put(39.5,43){\line(0,1){2}}

\put(17.5,-15){\qbezier(0,0)(0,8)(8,15.5)\qbezier(8,15.5)(16,23)(16,30)\qbezier(16,0)(16,8)(12.5,10.5)\qbezier(6,16)(0,23)(0,30)
\qbezier(6,0)(6,8)(14.5,15)\qbezier(14.5,15)(22,23)(22,30)\qbezier(22,0)(22,8)(16,13.5)\qbezier(9.5,19)(6,23)(6,30)}

\put(1.5,15){\qbezier(22,0)(22,8)(16,13.5)\qbezier(13,16)(6,23)(6,30)\qbezier(6,0)(6,8)(12,13)\qbezier(12,13)(22,23)(22,30)
\qbezier(16,0)(16,8)(12.5,11)\qbezier(9,14)(0,23)(0,30)\qbezier(0,0)(0,9)(6,14)\qbezier(12,19.5)(16,23)(16,30)}

\put(17.5,45){\qbezier(0,0)(0,8)(6,13.5)\qbezier(8.5,16)(16,23)(16,30)\qbezier(16,0)(16,8)(12.5,10.5)\qbezier(9.5,12.5)(0,23)(0,30)
\qbezier(6,0)(6,8)(13.5,14)\qbezier(15.5,16)(22,23)(22,30)\qbezier(22,0)(22,8)(12,17.5)\qbezier(9.5,19)(6,23)(6,30)}

\end{picture}
\caption{Two qubits with entangled strands.
\label{2ent0}}
\end{subfigure}%
 \begin{subfigure}{.35\textwidth}
\begin{picture}(200,190)(-50,-75)
\put(0,-50){
\put(0,0){\line(1,0){25}}
\put(0,0){\line(0,1){35}}
\put(25,35){\line(-1,0){25}}
\put(25,35){\line(0,-1){35}}
\put(5,13){\hbox{$\mathcal{B}^{(1)}_{1}$}}
\qbezier(3,0)(3,-8)(6,-8)
\qbezier(9,0)(9,-8)(6,-8)
\qbezier(16,0)(16,-8)(19,-8)
\qbezier(22,0)(22,-8)(19,-8)
}
\put(0,76){
\put(0,0){\line(1,0){25}}
\put(0,0){\line(0,1){35}}
\put(25,35){\line(-1,0){25}}
\put(25,35){\line(0,-1){35}}
\put(5,13){\hbox{$\mathcal{B}^{(1)}_{2}$}}
\qbezier(3,35)(3,43)(6,43)
\qbezier(9,35)(9,43)(6,43)
\qbezier(16,35)(16,43)(19,43)
\qbezier(22,35)(22,43)(19,43)
}
\put(36,-50){
\put(0,0){\line(1,0){25}}
\put(0,0){\line(0,1){35}}
\put(25,35){\line(-1,0){25}}
\put(25,35){\line(0,-1){35}}
\put(5,13){\hbox{$\mathcal{B}^{(2)}_1$}}
\qbezier(3,0)(3,-8)(6,-8)
\qbezier(9,0)(9,-8)(6,-8)
\qbezier(16,0)(16,-8)(19,-8)
\qbezier(22,0)(22,-8)(19,-8)
}
\put(36,76){
\put(0,0){\line(1,0){25}}
\put(0,0){\line(0,1){35}}
\put(25,35){\line(-1,0){25}}
\put(25,35){\line(0,-1){35}}
\put(5,13){\hbox{$\mathcal{B}^{(2)}_2$}}
\qbezier(3,35)(3,43)(6,43)
\qbezier(9,35)(9,43)(6,43)
\qbezier(16,35)(16,43)(19,43)
\qbezier(22,35)(22,43)(19,43)
}

\put(9,-15){\line(0,1){91}}
\put(3,-15){\line(0,1){91}}
\put(57,-15){\line(0,1){91}}
\put(51,-15){\line(0,1){91}}
\put(16,-15){\line(0,1){20}}
\put(22,-15){\line(0,1){20}}
\put(39,-15){\line(0,1){20}}
\put(45,-15){\line(0,1){20}}
\put(16,56){\line(0,1){20}}
\put(22,56){\line(0,1){20}}
\put(39,56){\line(0,1){20}}
\put(45,56){\line(0,1){20}}

\put(16,5){\qbezier(0,0)(0,13)(10,18)
\qbezier(16,21)(25,25)(15,30)
\qbezier(15,30)(0,35)(0,51)}
\put(22,5){\qbezier(0,0)(0,12)(8,15)
\qbezier(14,19)(25,25)(15,30)
\qbezier(15,30)(0,35)(0,51)}
\put(39,5){\qbezier(0,0)(0,15)(-15,20)
\qbezier(-15,20)(-25,25)(-14,31)
\qbezier(-8,35)(0,38)(0,51)}
\put(45,5){\qbezier(0,0)(0,15)(-15,20)
\qbezier(-15,20)(-25,25)(-16,29)
\qbezier(-10,32)(0,37)(0,51)}

\end{picture}
\caption{Two qubits entangled with cables.
}
\end{subfigure}
\caption{Different entanglements of two qubits. Most of them move the system out of computational space.\label{2ent1}}
\end{figure}
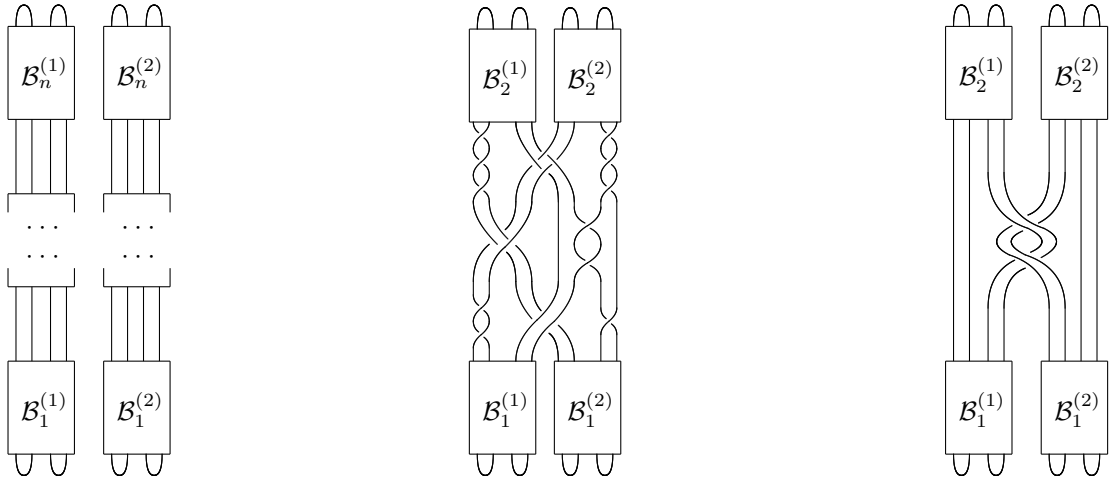

Two-qubit operations are always more complicated than one-qubit. They usually possess (much) lower fidelity and pose a major obstacle to building an efficient quantum computer. How can we build them for a topological quantum computer with effective Chern-Simons theory for generic $N$?

We want an operation which will combine two one-qubit structures and entangle them, see Fig.\ref{2ent1}. The main problem with such an approach is the dimension of the operators acting on such an 8-strand braid. If we take a product of four fundamental and four anti-fundamental representations, we will get 24 trivial representations rather than the four required for a two-qubit operation. This also leads to the problem that we do not know all the matrices acting in such a braid.

From the point of view of quantum computer such a two-qubit operation actually takes us from the product of two qubits to other states of the system, for example when each of the four-strand braid is in adjoint representation. We will say that the system is outside the computational space in this case. How can we make the two-qubit operation work as intended?

\begin{figure}[h!]
 \begin{subfigure}{.4\textwidth}
\begin{picture}(200,270)(-75,-75)
\put(0,-50){
\put(0,0){\line(1,0){25}}
\put(0,0){\line(0,1){35}}
\put(25,35){\line(-1,0){25}}
\put(25,35){\line(0,-1){35}}
\put(5,13){\hbox{$\mathcal{B}^{(1)}_{1}$}}
\put(3,0){\line(0,-1){5}}
\put(3,-8){\line(0,-1){3}}
\put(9,0){\line(0,-1){5}}
\put(9,-8){\line(0,-1){3}}
\put(16,0){\line(0,-1){5}}
\put(16,-8){\line(0,-1){3}}
\put(22,0){\line(0,-1){5}}
\put(22,-8){\line(0,-1){3}}
\put(3,-14){\line(0,-1){3}}
\put(9,-14){\line(0,-1){3}}
\put(16,-14){\line(0,-1){3}}
\put(22,-14){\line(0,-1){3}}
}
\put(0,105){
\put(0,0){\line(1,0){25}}
\put(0,0){\line(0,1){35}}
\put(25,35){\line(-1,0){25}}
\put(25,35){\line(0,-1){35}}
\put(5,13){\hbox{$\mathcal{B}^{(1)}_{2}$}}
\put(3,35){\line(0,1){5}}
\put(3,43){\line(0,1){3}}
\put(9,35){\line(0,1){5}}
\put(9,43){\line(0,1){3}}
\put(16,35){\line(0,1){5}}
\put(16,43){\line(0,1){3}}
\put(22,35){\line(0,1){5}}
\put(22,43){\line(0,1){3}}
\put(3,49){\line(0,1){3}}
\put(9,49){\line(0,1){3}}
\put(16,49){\line(0,1){3}}
\put(22,49){\line(0,1){3}}
}
\put(32,-50){
\put(0,0){\line(1,0){25}}
\put(0,0){\line(0,1){35}}
\put(25,35){\line(-1,0){25}}
\put(25,35){\line(0,-1){35}}
\put(5,13){\hbox{$\mathcal{B}^{(2)}_1$}}
\put(3,0){\line(0,-1){5}}
\put(3,-8){\line(0,-1){3}}
\put(9,0){\line(0,-1){5}}
\put(9,-8){\line(0,-1){3}}
\put(16,0){\line(0,-1){5}}
\put(16,-8){\line(0,-1){3}}
\put(22,0){\line(0,-1){5}}
\put(22,-8){\line(0,-1){3}}
\put(3,-14){\line(0,-1){3}}
\put(9,-14){\line(0,-1){3}}
\put(16,-14){\line(0,-1){3}}
\put(22,-14){\line(0,-1){3}}
}
\put(32,105){
\put(0,0){\line(1,0){25}}
\put(0,0){\line(0,1){35}}
\put(25,35){\line(-1,0){25}}
\put(25,35){\line(0,-1){35}}
\put(5,13){\hbox{$\mathcal{B}^{(2)}_2$}}
\put(3,35){\line(0,1){5}}
\put(3,43){\line(0,1){3}}
\put(9,35){\line(0,1){5}}
\put(9,43){\line(0,1){3}}
\put(16,35){\line(0,1){5}}
\put(16,43){\line(0,1){3}}
\put(22,35){\line(0,1){5}}
\put(22,43){\line(0,1){3}}
\put(3,49){\line(0,1){3}}
\put(9,49){\line(0,1){3}}
\put(16,49){\line(0,1){3}}
\put(22,49){\line(0,1){3}}
}

\green\put(1.5,-15){\line(0,1){30}}
\put(1.5,75){\line(0,1){30}}
\put(7.5,-15){\line(0,1){30}}
\put(7.5,75){\line(0,1){30}}
\black
\put(55.5,-15){\line(0,1){120}}
\put(49.5,-15){\line(0,1){120}}

{\red\put(33.5,15){\line(0,1){60}}\put(39.5,15){\line(0,1){60}}}

\put(17.5,-15){\red\qbezier(0,0)(0,8)(8,15.5)\qbezier(8,15.5)(16,23)(16,30)\blue\qbezier(16,0)(16,8)(12.5,10.5)\qbezier(6,16)(0,23)(0,30)
\red\qbezier(6,0)(6,8)(14.5,15)\qbezier(14.5,15)(22,23)(22,30)\blue\qbezier(22,0)(22,8)(16,13.5)\qbezier(9.5,19)(6,23)(6,30)}

\put(1.5,15){\blue\qbezier(22,0)(22,8)(14,15.5)\qbezier(14,15.5)(6,23)(6,30)\green\qbezier(6,0)(6,8)(9.5,10.5)\qbezier(16,16)(22,23)(22,30)
\blue\qbezier(16,0)(16,8)(7.5,15)\qbezier(7.5,15)(0,23)(0,30)\green\qbezier(0,0)(0,8)(6,13.5)\qbezier(12.5,19)(16,23)(16,30)}

\put(1.5,45){\green\qbezier(22,0)(22,8)(14,15.5)\qbezier(14,15.5)(6,23)(6,30)\blue\qbezier(6,0)(6,8)(9.5,10.5)\qbezier(16,16)(22,23)(22,30)
\green\qbezier(16,0)(16,8)(7.5,15)\qbezier(7.5,15)(0,23)(0,30)\blue\qbezier(0,0)(0,8)(6,13.5)\qbezier(12.5,19)(16,23)(16,30)}

\put(17.5,75){\blue\qbezier(0,0)(0,8)(8,15.5)\qbezier(8,15.5)(16,23)(16,30)\red\qbezier(16,0)(16,8)(12.5,10.5)\qbezier(6,16)(0,23)(0,30)
\blue\qbezier(6,0)(6,8)(14.5,15)\qbezier(14.5,15)(22,23)(22,30)\red\qbezier(22,0)(22,8)(16,13.5)\qbezier(9.5,19)(6,23)(6,30)}

\put(-20,-20){\line(1,0){100}}
\put(-20,110){\line(1,0){100}}
\put(-20,-20){\line(0,1){130}}
\put(80,-20){\line(0,1){130}}
\end{picture}
\end{subfigure}
\begin{subfigure}{.1\textwidth}
\begin{picture}(200,270)(0,-75)

\put(1.5,50){\line(1,0){30}}
\put(31.5,50){\line(-1,1){10}}
\put(31.5,50){\line(-1,-1){10}}
\end{picture}
\end{subfigure}
\begin{subfigure}{.5\textwidth}
\begin{picture}(200,270)(-75,-75)

\green\put(1.5,-15){\line(0,1){30}}
\put(1.5,75){\line(0,1){30}}
\black\put(50.5,-15){\line(0,1){120}}

\red\put(33.5,15){\line(0,1){60}}

\put(17.5,-15){\red\qbezier(0,0)(0,8)(8,15.5)\qbezier(8,15.5)(16,23)(16,30)\blue\qbezier(16,0)(16,8)(10.5,12.5)\qbezier(5,17)(0,23)(0,30)}

\put(1.5,15){\blue\qbezier(16,0)(16,8)(8.5,15)\qbezier(8.5,15)(0,23)(0,30)\green\qbezier(0,0)(0,8)(5.5,13)\qbezier(10.5,17)(16,23)(16,30)}

\put(1.5,45){\green\qbezier(16,0)(16,8)(8.5,15)\qbezier(8.5,15)(0,23)(0,30)\blue\qbezier(0,0)(0,8)(5.5,13)\qbezier(10.5,17)(16,23)(16,30)}

\put(17.5,75){\blue\qbezier(0,0)(0,8)(8,15.5)\qbezier(8,15.5)(16,23)(16,30)\red\qbezier(16,0)(16,8)(10.5,12.5)\qbezier(5,17)(0,23)(0,30)}
\put(0.4,0){\green\put(1.5,-15){\line(0,1){30}}
\put(1.5,75){\line(0,1){30}}
\black\put(50.5,-15){\line(0,1){120}}

\red\put(33.5,15){\line(0,1){60}}

\put(17.5,-15){\red\qbezier(0,0)(0,8)(8,15.5)\qbezier(8,15.5)(16,23)(16,30)\blue\qbezier(16,0)(16,8)(10.5,12.5)\qbezier(5,17)(0,23)(0,30)}

\put(1.5,15){\blue\qbezier(16,0)(16,8)(8.5,15)\qbezier(8.5,15)(0,23)(0,30)\green\qbezier(0,0)(0,8)(5.5,13)\qbezier(10.5,17)(16,23)(16,30)}

\put(1.5,45){\green\qbezier(16,0)(16,8)(8.5,15)\qbezier(8.5,15)(0,23)(0,30)\blue\qbezier(0,0)(0,8)(5.5,13)\qbezier(10.5,17)(16,23)(16,30)}

\put(17.5,75){\blue\qbezier(0,0)(0,8)(8,15.5)\qbezier(8,15.5)(16,23)(16,30)\red\qbezier(16,0)(16,8)(10.5,12.5)\qbezier(5,17)(0,23)(0,30)}
}
\end{picture}
\end{subfigure}
\caption{Entangling of two qubits with the cables on the left can be interpreted as a braid in higher representations on the right. These strands can be in any of the higher representations in the product of initial ones, leading to different configurations in (\ref{eq:config}).\label{2ent3}}
\end{figure}
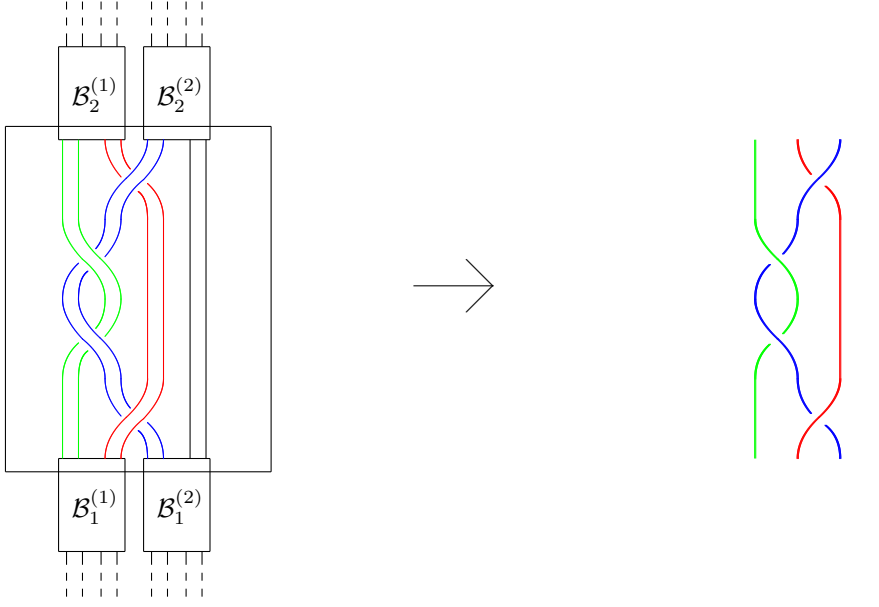

The main idea was presented in \cite{EntSU2}. We should take very specific entanglements that keep the system almost entirely within the computational space. There are three steps in this approach:

\textbf{1. Let us take cables instead of separate strands, see Fig.\ref{2ent3}.}

\textbf{2. Let us take such entanglements that each cable returns back to the place it started from.} It minimizes the probability leakage due to the different representations in cables in one qubit.

\textbf{3. Take entanglement with specifically chosen knot polynomials.} This is the main idea of the approach, so let us discuss it in more detail.

After replacing the strands with cables we reduce the initial configuration of the braid to four cases. The left and right halves of the braid are in the trivial representation. Therefore the representations of the left pair of cables coincide with each other; the same holds for the right pair.

Then the structure depends on whether the strands in each cable are co-directional or counter-directional. Let us discuss these cases separately.

\subsection{Co-directional cables}

If cables are co-directional then each cable can be in symmetric $[2]$ or antisymmetric $[1,1]$ representation:
\begin{equation}
[1]\otimes [1]= [2]+[1,1].
\end{equation}
Consequently initially we have four possible configurations:
\begin{equation}
\begin{array}{llclclcl}
\text{\textbf{I.}} & [2] & \otimes & [2] & \otimes & [2] & \otimes & [2]
\\
\text{\textbf{II.}} & [2] & \otimes & [2] & \otimes & [1,1] & \otimes & [1,1]
\\
\text{\textbf{III.}} & [1,1] & \otimes & [1,1] & \otimes & [2] & \otimes & [2]
\\
\text{\textbf{IV.}} & [1,1] & \otimes & [1,1] & \otimes & [1,1] & \otimes & [1,1]
\end{array}
\label{eq:config}
\end{equation}
Configurations cannot transform into one another, so in the end we remain within the same four sectors. However, for the system to be in the computational space, both the left and the right pair of cables should be in the trivial representation. This condition, however, need not be preserved during the intermediate evolution. It is not clear whether a non-trivial entanglement of cables can return the system to the computational space exactly. But we propose to try to find an entanglement which can achieve it with a probability close to $1$.

If we look at the resulting entanglement of the cables, we can see that it again looks like a $4$-plat braid. Therefore we can use the formulae from s.\ref{s:hrep} to get the matrices corresponding to the braid. And we are interested again in the element $\mathcal{B}_{\emptyset\emptyset}$ of the corresponding matrices, since in computational space both left and right pair of cables should be in trivial representation.

Since each of the four configurations can only transform to itself or move out of the computational space, we can say that the resulting two-qubit operation can be interpreted as a diagonal matrix of the size four by four:
\begin{equation}
\mathcal{O}=\left(
\begin{array}{cccc}
\mathcal{B}^{([2]\times[2])}_{\emptyset\emptyset}
\\
&
\mathcal{B}^{([2]\times[1,1])}_{\emptyset\emptyset}
\\
&&
\mathcal{B}^{([1,1]\times[2])}_{\emptyset\emptyset}
\\
&&&
\mathcal{B}^{([1,1]\times[1,1])}_{\emptyset\emptyset}
\end{array}\right),
\end{equation}
where elements correspond to the braids in corresponding representations.

The success probability of the operation in each configuration is the modulus squared of the corresponding element. The worst-case success probability is therefore the minimum of these values:
\begin{equation}
f_{\mathcal{O}}=\min_i\left|\mathcal{B}^{i}_{\emptyset\emptyset}\right|^2
\end{equation}
If we ignore leakage and keep only the phases, the operation is represented by a diagonal matrix with phase factors in each entry:
\begin{equation}
\mathcal{O}_p=\left(
\begin{array}{cccc}
\text{exp}\left(i\phi_{2,2}\right)
\\
&
\text{exp}\left(i\phi_{2,1}\right)
\\
&&
\text{exp}\left(i\phi_{1,2}\right)
\\
&&&
\text{exp}\left(i\phi_{1,1}\right)
\end{array}\right).
\end{equation}
An overall phase is irrelevant, and two one-qubit phase rotations can be absorbed; therefore the entangling content of this operation is described by
\begin{equation}
\mathcal{O}_{\text{eff}}=\left(
\begin{array}{cccc}
1
\\
&1
\\
&&1
\\
&&&
\text{exp}\left(i\left(\phi_{2,2}+\phi_{1,1}-\phi_{2,1}-\phi_{1,2}\right)\right)
\end{array}\right).
\label{eq:Oeff}
\end{equation}
The operation is a non-trivial entangling operation if the phase in the last element is non-trivial.

Therefore to get a non-trivial operation we need to get an entanglement where all the elements are as close to $1$ as possible and with the phase from (\ref{eq:Oeff}) to be non-trivial.

\subsection{Counter-directional cables}

If cables are counter-directional then each cable can be in the trivial $\emptyset$ or the adjoint representation:
\begin{equation}
[1]\otimes \overline{[1]}= \emptyset+\text{Adj}.
\end{equation}
Consequently initially we have four possible configurations:
\begin{equation}
\begin{array}{llclclcl}
\text{\textbf{I.}} & \emptyset & \otimes & \emptyset & \otimes & \emptyset & \otimes & \emptyset
\\
\text{\textbf{II.}} & \emptyset & \otimes & \emptyset & \otimes & \text{Adj} & \otimes & \text{Adj}
\\
\text{\textbf{III.}} & \text{Adj} & \otimes & \text{Adj} & \otimes & \emptyset & \otimes & \emptyset
\\
\text{\textbf{IV.}} & \text{Adj} & \otimes & \text{Adj} & \otimes & \text{Adj} & \otimes & \text{Adj}
\end{array}
\label{eq:config2}
\end{equation}
Repeating the same logic as in the co-directional case, we are interested again in the element $\mathcal{B}_{\emptyset\emptyset}$ of the corresponding braid matrix.

Since each of the four configurations can only transform to itself or move out of the computational space, we can say that the resulting two-qubit operation can be interpreted as a diagonal matrix of the size four by four:
\begin{equation}
\mathcal{O}=\left(
\begin{array}{cccc}
\mathcal{B}^{(\emptyset\times\emptyset)}_{\emptyset\emptyset}
\\
&
\mathcal{B}^{(\emptyset\times\text{Adj})}_{\emptyset\emptyset}
\\
&&
\mathcal{B}^{(\text{Adj}\times\emptyset)}_{\emptyset\emptyset}
\\
&&&
\mathcal{B}^{(\text{Adj}\times\text{Adj})}_{\emptyset\emptyset}
\end{array}\right),
\end{equation}
where elements correspond to the braids in corresponding representations.

The trivial representation braids trivially with everything, therefore
\begin{equation}
\mathcal{B}^{(\emptyset\times\emptyset)}_{\emptyset\emptyset}=1.
\end{equation}
Also, when we look at configurations $II$ and $III$, there is no non-trivial braiding between the trivial cables and the adjoint ones. Therefore the product of adjoint cables does not move away from the trivial representation, so
\begin{equation}
\left|\mathcal{B}^{(\emptyset\times\text{Adj})}_{\emptyset\emptyset}\right|^2=\left|\mathcal{B}^{( \text{Adj}\times\emptyset)}_{\emptyset\emptyset}\right|^2=1.
\end{equation}
The only non-trivial part of these elements is the phase, which comes from braiding the adjoint cables, and the corresponding eigenvalue is equal to $1/A^2$. Therefore in this case we need to find the entanglement for which $\mathcal{B}^{(\text{Adj}\times\text{Adj})}_{\emptyset\emptyset}$ is close to $1$ with the whole phase being non-trivial.

The success probability of the operation is again the modulus squared of the corresponding element. In the present counter-directional setup, the only non-trivial contribution comes from configuration $IV$, so the success probability is:
\begin{equation}
f_{\mathcal{O}}=
\left|\mathcal{B}^{(\text{Adj}\times\text{Adj})}_{\emptyset\emptyset}\right|^2
\end{equation}
If we again ignore leakage and keep only the phases, the operation is a diagonal matrix with phase factors:
\begin{equation}
\mathcal{O}_p=\left(
\begin{array}{cccc}
1
\\
&
\text{exp}\left(i\phi_{0,a}\right)
\\
&&
\text{exp}\left(i\phi_{a,0}\right)
\\
&&&
\text{exp}\left(i\phi_{a,a}\right)
\end{array}\right),
\end{equation}
which again corresponds to the following
\begin{equation}
\mathcal{O}_{\text{eff}}=\left(
\begin{array}{cccc}
1
\\
&1
\\
&&1
\\
&&&
\text{exp}\left(i\left(\phi_{a,a}-\phi_{0,a}-\phi_{a,0}\right)\right)
\end{array}\right).
\label{eq:Oeff2}
\end{equation}

\section{Framing of $\mathcal{R}$-matrices\label{s:fr}}

In section \ref{s:Rm} we discussed what is the form of quantum $\mathcal{R}$-matrices. However, by definition these matrices are the solutions to the Yang-Baxter equation:
\begin{equation}
\mathcal{R}_1\mathcal{R}_2\mathcal{R}_1=\mathcal{R}_2\mathcal{R}_1\mathcal{R}_2.
\end{equation}
It is easy to see that if we multiply the $\mathcal{R}$-matrix by some coefficient, the equation still holds. This poses the question: what is the ``correct'' normalization of the $\mathcal{R}$-matrix?

The answer is -- it depends. From the representation theory of quantum groups \cite{Klimyk} we know that eigenvalues for some representation $Y$ are proportional to
\begin{equation}
\lambda_Y\sim q^{a_Y},\ \ \ a_Y=(v_Y,v_Y+2\rho),
\end{equation}
where $v_Y$ is the highest weight vector of the representation $Y$, and $\rho$ is a half-sum of positive roots of the algebra. If we calculate this product directly for the group $SU(N)$ and a Young diagram $Y$, we get
\begin{equation}
a_Y=\varkappa_Y+\frac{N|Y|}{2}-\frac{|Y|^2}{2N}.
\end{equation}
This is, however, not the whole story. The exact normalization, called framing in knot theory \cite{fr1,fr2,fr3,Cable}, depends on the problem in which we use the $\mathcal{R}$-matrix. In knot theory we are usually interested in topologically invariant answers for the polynomials. This leads to the existence of the first Reidemeister move, which includes only one $\mathcal{R}$-matrix and, therefore, can be used to fix the normalization.

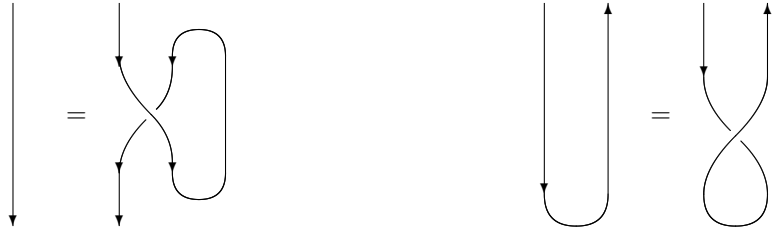
\begin{figure}[h!]
\begin{picture}(200,100)(-100,-100)
\put(0,0){
\put(0,0){\vector(0,-1){84}}
\put(20,-45){\hbox{$=$}}
\put(40,0){\line(0,-1){20}}
\qbezier(40,-20)(40,-30)(52,-42)
\qbezier(40,-64)(40,-54)(50,-44)
\put(40,-64){\vector(0,-1){20}}
\put(40,-20){\vector(0,-1){4}}
\put(40,-60){\vector(0,-1){4}}
\qbezier(52,-42)(60,-50)(60,-60)
\qbezier(54,-40)(60,-34)(60,-24)
\put(60,-20){\vector(0,-1){4}}
\put(60,-60){\vector(0,-1){4}}
\qbezier(60,-20)(60,-10)(70,-10)
\qbezier(80,-20)(80,-10)(70,-10)
\qbezier(60,-64)(60,-74)(70,-74)
\qbezier(80,-64)(80,-74)(70,-74)
\put(80,-20){\line(0,-1){44}}
}
\put(200,0){
\put(0,0){\vector(0,-1){72}}
\put(24,-72){\vector(0,1){72}}
\qbezier(0,-72)(0,-84)(12,-84)
\qbezier(24,-72)(24,-84)(12,-84)
\put(40,-45){\hbox{$=$}}
\put(10,0){
\put(50,0){\vector(0,-1){28}}
\put(74,-28){\vector(0,1){28}}
\qbezier(50,-72)(50,-62)(62,-50)
\qbezier(74,-28)(74,-38)(62,-50)
\qbezier(74,-72)(74,-62)(64,-52)
\qbezier(50,-28)(50,-38)(60,-48)
\qbezier(50,-72)(50,-84)(62,-84)
\qbezier(74,-72)(74,-84)(62,-84)
}
}
\end{picture}
\caption{Two types of the first Reidemeister move for parallel (on the left) and anti-parallel (on the right) strands.\label{f:1Reid}}
\end{figure}

The first Reidemeister move exists only when two crossing strands belong to the same component and depends on the type of the crossing, see Fig. \ref{f:1Reid}. In the first case when two parallel strands cross the correct topological framing is equal to
\begin{equation}
|\lambda_Q|=q^{a_Q-4a_Y}, \text{ if } Q\in Y\otimes Y.
\end{equation}
In the second case, when two anti-parallel strands cross the topological framing is
\begin{equation}
|\lambda_Q|=q^{a_Q}, \text{ if } Q\in Y\otimes \overline{Y}.
\end{equation}
If we have a crossing between two different components, then the first Reidemeister move does not exist. Basically, algebraic number of such crossings is a link invariant - it cannot be changed without changing the topology of the knot. Therefore this normalization can be chosen arbitrarily from the point of view of topological invariance.

There is, however, a different consideration, especially relevant to the discussion in this paper. If we talk about cabling, then we expect that the matrix in higher representations can be expressed through matrices in smaller representations. And this property again depends on the choice of normalization for all the matrices. This connection holds if we choose eigenvalues in topological framing \cite{Cable}:
\begin{equation}
|\lambda_Q|=q^{a_Q-a_{Y_1}-a_{Y_2}}, \text{ if } Q\in Y_1\otimes Y_2.
\label{eq:vfr}
\end{equation}

If we speak about a topological quantum computer, the general choice of framing in fact does not matter, since the relevant observable -- the process probability -- is the modulus squared of the matrix element corresponding to the entangled braid. And all normalizations of the $\mathcal{R}$-matrix differ only by phase factor. However, when we use our algorithm for the two-qubit operation, it is important that the phases in different configurations are mutually consistent.

Let us recall that, according to the discussed model, the qubit topological quantum computer is based on anyons in the fundamental representation. The higher-representation structures used to build the two-qubit operation are in fact constructed from them. Therefore it is logical to consider all the matrices in vertical framing (\ref{eq:vfr}). And all the matrices in s.\ref{s:Rm} are given in this framing.

\section{Examples}

For several arbitrary values of $k$ and $N$, we found candidate entanglements with the required properties using brute force. For the co-directional cables the answers we found are:

\begin{equation*}
\begin{array}{|l|l|c|l|l|}
\hline
N & k & \text{Braid} & \text{Presicion} & \Delta\phi
\\
\hline
3 & 30&  [2, -4, 2, -4, 2, 8, 2] & 0.98 & 0.25
\\
\hline
3 & 35&  [2, 6, 2, -10, 2, 6, 2] & 0.98 & 0.31
\\
\hline
4 & 24&  [4, -10, -14, -4, 10] & 1 & 2
\\
\hline
4 & 29&  [2, -2, -16, 2, -2] & 0.98 & 2.01
\\
\hline
5 & 30&  [2, 8, 8, -6, -10, -10, -2] & 0.98 & 1.96
\\
\hline
5 & 35&  [2, -8, 18, -2, 10] & 0.98 & 2.01
\\
\hline
5 & 35&  [2, -8, 18, -2, 10] & 0.98 & 2.01
\\
\hline
6 & 39&  [2, -16, -2, 6, -10] & 0.98 & 2.01
\\
\hline
\end{array}
\end{equation*}

For the counter-directional cables we found the following combinations:

\begin{equation*}
\begin{array}{|l|l|c|l|l|}
\hline
N & k & \text{Braid} & \text{Precision} & \Delta\phi
\\
\hline
3 & 30& [2, -6, -2, -6, 2, -6, -2] &  0.997 & 2.66
\\
\hline
4 & 41 &  [10, -16, -22, -6, -10] & 0.995 & 0.24
\\
\hline
4 & 49&  [2, -8, -2, 2, 2, -8, -2]  & 0.993 & 0.38
\\
\hline
5 & 49& [2, -10, -2, -6, 2, -10, -2] &  0.994 & 0.20
\\
\hline
5 & 54 & [2, -4, 2, -2, 2, -4, 2] & 0.996 & 0.13
\\
\hline
\end{array}
\end{equation*}

\begin{figure}[h!]
\begin{picture}(200,150)(-170,-150)
\put(0,0){\line(0,-1){40}}
\put(24,0){\line(0,-1){10}}
\put(48,0){\line(0,-1){10}}
\put(72,0){\line(0,-1){120}}
\qbezier(24,-10)(24,-20)(29,-25)
\qbezier(48,-10)(48,-20)(43,-25)
\put(29,-20){
\put(0,0){\line(1,0){14}}
\put(0,0){\line(0,-1){30}}
\put(0,-30){\line(1,0){14}}
\put(14,0){\line(0,-1){30}}
\put(3,-17){\hbox{$n_1$}}
}
\qbezier(43,-45)(48,-50)(48,-60)
\put(29,-45){\line(-1,-1){10}}
\qbezier(0,-40)(0,-50)(5,-55)
\put(5,-50){
\put(0,0){\line(1,0){14}}
\put(0,0){\line(0,-1){30}}
\put(0,-30){\line(1,0){14}}
\put(14,0){\line(0,-1){30}}
\put(3,-17){\hbox{$n_2$}}
}
\qbezier(0,-90)(0,-80)(5,-75)
\put(19,-75){\line(1,-1){10}}
\put(48,-60){\line(0,-1){10}}
\qbezier(48,-70)(48,-80)(43,-85)
\put(29,-80){
\put(0,0){\line(1,0){14}}
\put(0,0){\line(0,-1){30}}
\put(0,-30){\line(1,0){14}}
\put(14,0){\line(0,-1){30}}
\put(3,-17){\hbox{$n_3$}}
}
\qbezier(29,-105)(24,-110)(24,-120)
\qbezier(43,-105)(48,-110)(48,-120)
\put(0,-90){\line(0,-1){30}}
\multiput(0,-122)(0,-4){5}{\line(0,-1){2}}
\multiput(24,-122)(0,-4){5}{\line(0,-1){2}}
\multiput(48,-122)(0,-4){5}{\line(0,-1){2}}
\multiput(72,-122)(0,-4){5}{\line(0,-1){2}}
\end{picture}
\caption{Notation for an entangling cabled braid $[n_1,n_2,n_3...]$, boxes denote repeated crossings.\label{f:brnote}}
\end{figure}
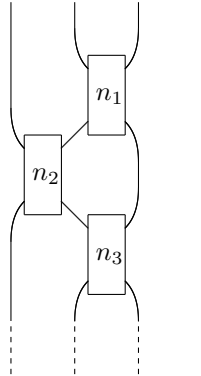

Here we use the notation for an entangling cabled braid illustrated by Fig. \ref{f:brnote}

The expanded braids for the numerical patterns from the table are shown on Figs. \ref{Brco} and \ref{Brcount}.
\newcommand{\BraidPosCenter}{%
\put(0,0){\qbezier(0.5,0)(0.5,3)(5.5,6)\qbezier(5.5,6)(9.5,9)(9.5,12)\qbezier(-0.5,0)(-0.5,3)(3.5,6)\qbezier(3.5,6)(8.5,9)(8.5,12)
\qbezier(9.5,0)(9.5,3)(6.5,5)\qbezier(3.5,7.5)(0.5,9)(0.5,12)\qbezier(8.5,0)(8.5,3)(5.5,4.5)\qbezier(2.5,7)(-0.5,9)(-0.5,12)
\put(18,0){\put(-0.5,0){\line(0,1){12}}\put(0.5,0){\line(0,1){12}}}\put(-18,0){\put(8.5,0){\line(0,1){12}}\put(9.5,0){\line(0,1){12}}}}
}
\newcommand{\BraidNegCenter}{%
\put(0,0){\qbezier(0.5,0)(0.5,3)(3.5,4.5)\qbezier(6.5,7)(9.5,9)(9.5,12)\qbezier(-0.5,0)(-0.5,3)(2.5,5)\qbezier(5.5,7.5)(8.5,9)(8.5,12)
\qbezier(8.5,0)(8.5,3)(3.5,6)\qbezier(3.5,6)(-0.5,9)(-0.5,12)\qbezier(9.5,0)(9.5,3)(5.5,6)\qbezier(5.5,6)(0.5,9)(0.5,12)
\put(18,0){\put(-0.5,0){\line(0,1){12}}\put(0.5,0){\line(0,1){12}}}\put(-18,0){\put(8.5,0){\line(0,1){12}}\put(9.5,0){\line(0,1){12}}}}
}
\newcommand{\BraidPosLeft}{%
\put(-9,0){\qbezier(0.5,0)(0.5,3)(5.5,6)\qbezier(5.5,6)(9.5,9)(9.5,12)\qbezier(-0.5,0)(-0.5,3)(3.5,6)\qbezier(3.5,6)(8.5,9)(8.5,12)
\qbezier(9.5,0)(9.5,3)(6.5,5)\qbezier(3.5,7.5)(0.5,9)(0.5,12)\qbezier(8.5,0)(8.5,3)(5.5,4.5)\qbezier(2.5,7)(-0.5,9)(-0.5,12)}
\put(9,0){\put(-0.5,0){\line(0,1){12}}\put(0.5,0){\line(0,1){12}}}\put(9,0){\put(8.5,0){\line(0,1){12}}\put(9.5,0){\line(0,1){12}}}
}
\newcommand{\BraidNegLeft}{%
\put(-9,0){\qbezier(0.5,0)(0.5,3)(3.5,4.5)\qbezier(6.5,7)(9.5,9)(9.5,12)\qbezier(-0.5,0)(-0.5,3)(2.5,5)\qbezier(5.5,7.5)(8.5,9)(8.5,12)
\qbezier(8.5,0)(8.5,3)(3.5,6)\qbezier(3.5,6)(-0.5,9)(-0.5,12)\qbezier(9.5,0)(9.5,3)(5.5,6)\qbezier(5.5,6)(0.5,9)(0.5,12)}
\put(9,0){\put(-0.5,0){\line(0,1){12}}\put(0.5,0){\line(0,1){12}}}\put(9,0){\put(8.5,0){\line(0,1){12}}\put(9.5,0){\line(0,1){12}}}
}
\begin{figure}[h!]
\begin{subfigure}{.3\textwidth}
\begin{picture}(0,400)(0,0)
\put(60,354){\makebox(0,0){$(N,k)=(3,30)$}}
\put(60,56){\BraidPosCenter}
\put(60,68){\BraidPosCenter}
\put(60,80){\BraidNegLeft}
\put(60,92){\BraidNegLeft}
\put(60,104){\BraidNegLeft}
\put(60,116){\BraidNegLeft}
\put(60,128){\BraidPosCenter}
\put(60,140){\BraidPosCenter}
\put(60,152){\BraidNegLeft}
\put(60,164){\BraidNegLeft}
\put(60,176){\BraidNegLeft}
\put(60,188){\BraidNegLeft}
\put(60,200){\BraidPosCenter}
\put(60,212){\BraidPosCenter}
\put(60,224){\BraidPosLeft}
\put(60,236){\BraidPosLeft}
\put(60,248){\BraidPosLeft}
\put(60,260){\BraidPosLeft}
\put(60,272){\BraidPosLeft}
\put(60,284){\BraidPosLeft}
\put(60,296){\BraidPosLeft}
\put(60,308){\BraidPosLeft}
\put(60,320){\BraidPosCenter}
\put(60,332){\BraidPosCenter}
\end{picture}
\end{subfigure}
\begin{subfigure}{.3\textwidth}
\begin{picture}(0,500)(0,0)
\put(60,442){\makebox(0,0){$(N,k)=(3,35)$}}
\put(60,72){\BraidPosCenter}
\put(60,84){\BraidPosCenter}
\put(60,96){\BraidPosLeft}
\put(60,108){\BraidPosLeft}
\put(60,120){\BraidPosLeft}
\put(60,132){\BraidPosLeft}
\put(60,144){\BraidPosLeft}
\put(60,156){\BraidPosLeft}
\put(60,168){\BraidPosCenter}
\put(60,180){\BraidPosCenter}
\put(60,192){\BraidNegLeft}
\put(60,204){\BraidNegLeft}
\put(60,216){\BraidNegLeft}
\put(60,228){\BraidNegLeft}
\put(60,240){\BraidNegLeft}
\put(60,252){\BraidNegLeft}
\put(60,264){\BraidNegLeft}
\put(60,276){\BraidNegLeft}
\put(60,288){\BraidNegLeft}
\put(60,300){\BraidNegLeft}
\put(60,312){\BraidPosCenter}
\put(60,324){\BraidPosCenter}
\put(60,336){\BraidPosLeft}
\put(60,348){\BraidPosLeft}
\put(60,360){\BraidPosLeft}
\put(60,372){\BraidPosLeft}
\put(60,384){\BraidPosLeft}
\put(60,396){\BraidPosLeft}
\put(60,408){\BraidPosCenter}
\put(60,420){\BraidPosCenter}
\end{picture}
\end{subfigure}
\begin{subfigure}{.3\textwidth}
\begin{picture}(0,316)(0,0)
\put(60,308){\makebox(0,0){$ (N,k)=(4,29)$}}
\put(60,0){\BraidPosCenter}
\put(60,12){\BraidPosCenter}
\put(60,24){\BraidNegLeft}
\put(60,36){\BraidNegLeft}
\put(60,48){\BraidNegCenter}
\put(60,60){\BraidNegCenter}
\put(60,72){\BraidNegCenter}
\put(60,84){\BraidNegCenter}
\put(60,96){\BraidNegCenter}
\put(60,108){\BraidNegCenter}
\put(60,120){\BraidNegCenter}
\put(60,132){\BraidNegCenter}
\put(60,144){\BraidNegCenter}
\put(60,156){\BraidNegCenter}
\put(60,168){\BraidNegCenter}
\put(60,180){\BraidNegCenter}
\put(60,192){\BraidNegCenter}
\put(60,204){\BraidNegCenter}
\put(60,216){\BraidNegCenter}
\put(60,228){\BraidNegCenter}
\put(60,240){\BraidPosLeft}
\put(60,252){\BraidPosLeft}
\put(60,264){\BraidNegCenter}
\put(60,276){\BraidNegCenter}
\end{picture}
\end{subfigure}
\caption{Examples of braid for the patterns from the table for co-directional cables.\label{Brco}}
\end{figure}
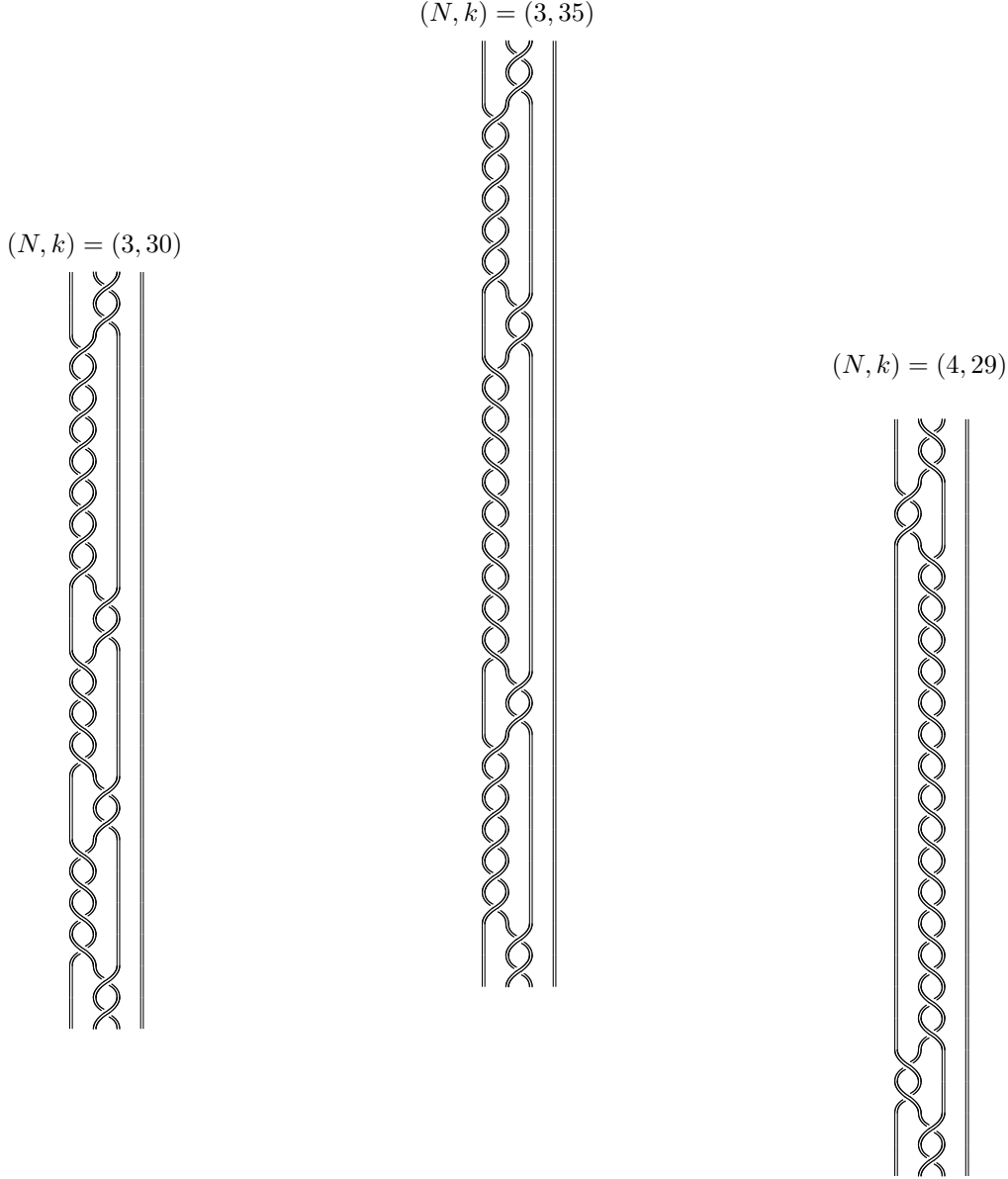
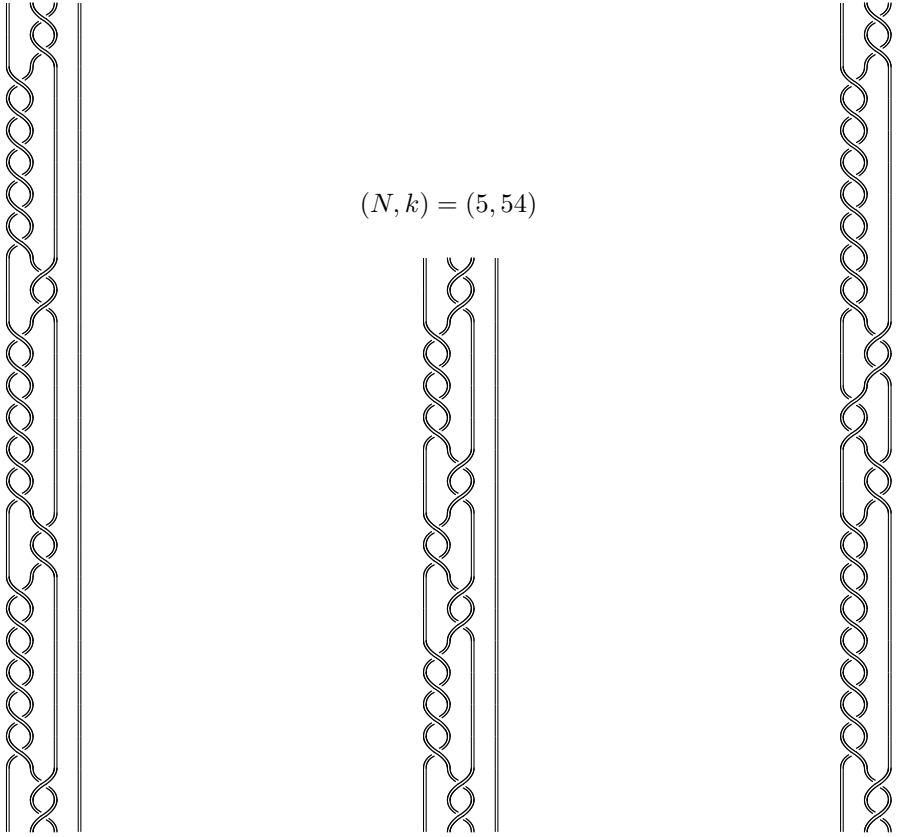
\begin{figure}[h!]
\begin{subfigure}{.3\textwidth}
\begin{picture}(0,340)(0,0)
\put(60,332){\makebox(0,0){$ (N,k)=(3,30)$}}
\put(60,0){\BraidPosCenter}
\put(60,12){\BraidPosCenter}
\put(60,24){\BraidNegLeft}
\put(60,36){\BraidNegLeft}
\put(60,48){\BraidNegLeft}
\put(60,60){\BraidNegLeft}
\put(60,72){\BraidNegLeft}
\put(60,84){\BraidNegLeft}
\put(60,96){\BraidNegCenter}
\put(60,108){\BraidNegCenter}
\put(60,120){\BraidNegLeft}
\put(60,132){\BraidNegLeft}
\put(60,144){\BraidNegLeft}
\put(60,156){\BraidNegLeft}
\put(60,168){\BraidNegLeft}
\put(60,180){\BraidNegLeft}
\put(60,192){\BraidPosCenter}
\put(60,204){\BraidPosCenter}
\put(60,216){\BraidNegLeft}
\put(60,228){\BraidNegLeft}
\put(60,240){\BraidNegLeft}
\put(60,252){\BraidNegLeft}
\put(60,264){\BraidNegLeft}
\put(60,276){\BraidNegLeft}
\put(60,288){\BraidNegCenter}
\put(60,300){\BraidNegCenter}
\end{picture}
\end{subfigure}
\begin{subfigure}{.3\textwidth}
\begin{picture}(0,244)(0,0)
\put(60,236){\makebox(0,0){$ (N,k)=(5,54)$}}
\put(60,0){\BraidPosCenter}
\put(60,12){\BraidPosCenter}
\put(60,24){\BraidNegLeft}
\put(60,36){\BraidNegLeft}
\put(60,48){\BraidNegLeft}
\put(60,60){\BraidNegLeft}
\put(60,72){\BraidPosCenter}
\put(60,84){\BraidPosCenter}
\put(60,96){\BraidNegLeft}
\put(60,108){\BraidNegLeft}
\put(60,120){\BraidPosCenter}
\put(60,132){\BraidPosCenter}
\put(60,144){\BraidNegLeft}
\put(60,156){\BraidNegLeft}
\put(60,168){\BraidNegLeft}
\put(60,180){\BraidNegLeft}
\put(60,192){\BraidPosCenter}
\put(60,204){\BraidPosCenter}
\end{picture}
\end{subfigure}
\begin{subfigure}{.3\textwidth}
\begin{picture}(0,340)(0,0)
\put(60,332){\makebox(0,0){$ (N,k)=(4,49)$}}
\put(60,0){\BraidPosCenter}
\put(60,12){\BraidPosCenter}
\put(60,24){\BraidNegLeft}
\put(60,36){\BraidNegLeft}
\put(60,48){\BraidNegLeft}
\put(60,60){\BraidNegLeft}
\put(60,72){\BraidNegLeft}
\put(60,84){\BraidNegLeft}
\put(60,96){\BraidNegLeft}
\put(60,108){\BraidNegLeft}
\put(60,120){\BraidNegCenter}
\put(60,132){\BraidNegCenter}
\put(60,144){\BraidPosLeft}
\put(60,156){\BraidPosLeft}
\put(60,168){\BraidPosCenter}
\put(60,180){\BraidPosCenter}
\put(60,192){\BraidNegLeft}
\put(60,204){\BraidNegLeft}
\put(60,216){\BraidNegLeft}
\put(60,228){\BraidNegLeft}
\put(60,240){\BraidNegLeft}
\put(60,252){\BraidNegLeft}
\put(60,264){\BraidNegLeft}
\put(60,276){\BraidNegLeft}
\put(60,288){\BraidNegCenter}
\put(60,300){\BraidNegCenter}
\end{picture}
\end{subfigure}
\caption{Examples of braid for the patterns from the table for counter-directional cables.\label{Brcount}}
\end{figure}

\section{Conclusion}

We showed how we can build two-qubit operations for a topological quantum computer based on Chern-Simons theory with the $SU(N)$ gauge group. This paper is the continuation of our idea, suggested in \cite{EntSU2} for $SU(2)$ group. To build effective high-fidelity two-qubit operations we suggested constructing a very specific entanglement of one-qubit braids. To find such entanglements we used cabling approach \cite{Cable}. This allows us to calculate operations corresponding to braids with many strands more easily, which is a difficult problem for generic entangled braids.

In $SU(N)$ case new problems arise -- in the $SU(2)$ case only one element in the resulting two-qubit operation was non-trivial and it was required to find a link, which polynomial is close to $1$ by modulo only in symmetric representation. In this case, however, the required link should have this property simultaneously for four different combination of representations. However, we managed to find such combinations for several arbitrarily chosen parameters of the theory, which shows that this is in fact not a big problem.

There are of course further questions one can ask. The first question is can we find required entanglements in anlytical rather than numerical way, can find some structure in such entanglements? The second question is can the same thing be done if we look at qudit rather than qubit computers. It means that we should take higher representations as a basic ones. It is known that qudit computers can be more powerful and have higher fidelity \cite{Kiktenko:2023ytz}, therefore, it is interesting if the same procedure can be applied there.

\section*{Acknowledgements}


This work was supported by the Russian Science Foundation grant No 23-71-10058.

\end{document}